\documentclass{emulateapj}

\newcommand{\Kepler}{{\it Kepler}}

\newcommand{\hipparcos}{{\it Hipparcos}}

\newcommand{\Gaia}{{\it Gaia}}
\newcommand{\be}{\begin{equation}}
\newcommand{\ee}{\end{equation}}
\newcommand{\ron}{\color{black}}

\newcommand{\metallicity}{[M/H]}
\newcommand{\msun}{M$_\odot$}

\newcommand{\kms}{\ensuremath{\rm km\,s^{-1}}}
\newcommand{\ms}{\ensuremath{\rm m\,s^{-1}}}
\newcommand{\msy}{\ensuremath{\rm m\,s^{-1}\,yr^{-1}}}

\newcommand{\loggspc}{4.7}
\newcommand{\loggespc}{0.1}

\newcommand{\mh}{0.15}
\newcommand{\mhe}{0.03}

\newcommand{\mstar}{0.74}
\newcommand{\mstare}{0.03}

\newcommand{\rstar}{0.66}
\newcommand{\rstare}{0.03}

\newcommand{\mearth}{M$_\oplus$}
\newcommand{\rearth}{R$_\oplus$}

\newcommand{\Alopeke}{`$\!$Alopeke}

\newcommand{\ldone}{0.570}
\newcommand{\uldone}{0.062}
\newcommand{\ldtwo}{0.043}
\newcommand{\uldtwo}{0.068}
\newcommand{\rprstb}{0.0272}
\newcommand{\urprstb}{0.0012}
\newcommand{\arstb}{129}
\newcommand{\uarstb}{^{+55}_{-22}}
\newcommand{\inclb}{89.744}
\newcommand{\uinclb}{^{+0.131}_{-0.085}}
\newcommand{\impb}{0.64}
\newcommand{\uimpb}{^{+0.13}_{-0.31}}
\newcommand{\rplb}{1.96}
\newcommand{\urplb}{0.13}
\newcommand{\perplb}{106}
\newcommand{\uperplb}{^{+74}_{-25}}
\newcommand{\ttransitb}{2457828.3869}
\newcommand{\uttransitb}{0.0011}
\newcommand{\teqb}{255}
\newcommand{\uteqb}{^{+38}_{-44}}
\newcommand{\tdurb}{4.600}
\newcommand{\utdurb}{0.097}

\newcommand{\thisstar}{HD~283869}
\newcommand{\thisplanet}{HD~283869~b}

\usepackage{xcolor}
\usepackage{hyperref}
\usepackage{breakurl}
\usepackage{amsmath}
\usepackage{graphicx}
\usepackage{verbatim}
\usepackage{booktabs}

\slugcomment{}

\shorttitle{A Single Transit in the Hyades}
\shortauthors{Vanderburg et al.}

\newcommand\ut{1}
\newcommand\col{3}
\newcommand\har{2}
\newcommand\noao{4}
\newcommand\ctc{5}
\newcommand\ames{6}
\newcommand\dtu{7}
\newcommand\ku{8}
\newcommand\sagan{9}

\newcommand\hub{10}

\newcommand\ema{11}

\begin{document}



\title{Zodiacal Exoplanets in Time (ZEIT) VII:\\ A Temperate Candidate Super-Earth in the Hyades Cluster}

\author{Andrew Vanderburg\altaffilmark{\ut,\har,\sagan,\ema}, Andrew W. Mann\altaffilmark{\col,\ut,\hub}, Aaron Rizzuto\altaffilmark{\ut}, Allyson Bieryla\altaffilmark{\har}, Adam L. Kraus\altaffilmark{\ut}, Perry Berlind\altaffilmark{\har}, Michael L. Calkins\altaffilmark{\har}, Jason L. Curtis\altaffilmark{\col}, Stephanie T. Douglas\altaffilmark{\har}, Gilbert A. Esquerdo\altaffilmark{\har}, Mark E. Everett\altaffilmark{\noao}, Elliott P. Horch\altaffilmark{\ctc}, Steve B. Howell\altaffilmark{\ames}, David W. Latham\altaffilmark{\har}, Andrew W. Mayo\altaffilmark{\dtu,\ku}, Samuel N. Quinn\altaffilmark{\har}, Nicholas J. Scott\altaffilmark{\ames}, Robert P. Stefanik\altaffilmark{\har}}
%

\altaffiltext{\ut}{Department of Astronomy, The University of Texas at Austin, Austin, TX 78712, USA}
\altaffiltext{\har}{Harvard--Smithsonian Center for Astrophysics, Cambridge, Massachusetts 02138, USA}
\altaffiltext{\col}{Department of Astronomy, Columbia University, 550 West 120th Street, New York, NY 10027, USA}
\altaffiltext{\noao}{National Optical Astronomy Observatory, 950 North Cherry Avenue, Tucson, AZ 85719}
\altaffiltext{\ctc}{Department of Physics, Southern Connecticut State University, 501 Crescent Street, New Haven, CT 06515}
\altaffiltext{\ames}{Space Science and Astrobiology Division, NASA Ames Research Center, Moffett Field, CA 94035}
\altaffiltext{\dtu}{DTU Space, National Space Institute, Technical University of Denmark, Elektrovej 327, DK-2800 Lyngby, Denmark}
\altaffiltext{\ku}{Centre for Star and Planet Formation, Natural History Museum of Denmark \& Niels Bohr Institute, University of Copenhagen, \O ster Voldgade 5-7, DK-1350 Copenhagen K.}
\altaffiltext{\sagan}{NASA Sagan Fellow}
\altaffiltext{\hub}{NASA Hubble Fellow}
\altaffiltext{\ema}{\url{avanderburg@utexas.edu}}


\begin{abstract}

Transiting exoplanets in young open clusters present opportunities to study how exoplanets evolve over their lifetimes. Recently, significant progress detecting transiting planets in young open clusters has been made with the K2 mission, but so far all of these transiting cluster planets orbit close to their host stars, so planet evolution can only be studied in a high-irradiation regime. Here, we report the discovery of a long-period planet candidate, called HD 283869 b, orbiting a member of the Hyades cluster. Using data from the K2 mission, we detected a single transit of a super-Earth-sized (1.96 $\pm$ 0.12 \rearth) planet candidate orbiting the K-dwarf HD 283869 with a period longer than 72 days. Since we only detected a single transit event, we cannot validate HD 283869 b with high confidence, but our analysis of the K2 images, archival data, and follow-up observations suggests that the source of the event is indeed a transiting planet. We estimated the candidate's orbital parameters and find that if real, it has a period P$\approx$100 days and receives approximately Earth-like incident flux, giving the candidate a 71\% chance of falling within the circumstellar habitable zone. If confirmed, HD 283869 b would have the longest orbital period, lowest incident flux, and brightest host star of any known transiting planet in an open cluster, making it uniquely important to future studies of how stellar irradiation affects planetary evolution. 

\end{abstract}


\section{Introduction}

The study of stars in clusters has been a cornerstone of stellar astrophysics for over a century \citep[e.g.][]{russell, shapley}. Because clusters contain coeval stellar populations with uniform ages, compositions and formation histories, it is possible to study stars while controlling for these variables, determine how stars of different masses appear and evolve, and understand cases where stellar evolution took unconventional paths. Stars in open clusters have enabled studies of, among other phenomena, stellar mergers \citep{leiner}, mass transfer \citep{geller}, rotation \citep{barnes}, and magnetic activity \citep{stern}. 


Now that in the last few decades, the detection of exoplanets has gone from unproven \citep{struve, campbellwalker1} to achievable \citep{campbellwalker2, lathambd, mayor, butler, cochran97}, to routine \citep{rowe, morton16, mayo}, fundamental questions about formation and evolution of exoplanets are becoming pertinent. Since the very first discoveries, exoplanets have been found with orbits \citep{mayor, hd80606, cochran97}, and interior structures/compositions \citep{charbonneau, masuda51} different from those of our own Solar System planets, in tension with traditional planet formation theories \citep[e.g.][]{boss}. As the number of detected exoplanets grows, increasingly sophisticated analyses are beginning to yield insights into these surprising features of the exoplanet population  \citep[e.g.][]{rogers, dawson}. 

As astronomers begin to tackle fundamental questions about the origin and evolution of exoplanets, it stands to reason that the study of exoplanets in clusters may be similarly foundational to the study of stars in clusters. Studying a coeval planet population within a cluster could isolate trends in planet properties as a function of stellar mass \citep{cochran}, while comparisons between different clusters and field populations could reveal how planet demographics depend on birth environment and how they change over time \citep{meibom, mann16}.

Recently, significant progress has been made detecting exoplanets in clusters. Some of the earliest discoveries came from radial velocity (RV) searches of cluster members \citep{sato, lovismayor, twobs} which were generally only sensitive to giant planets. Searches for transits were originally unfruitful \citep{gilliland, burkecluster, pepper}\footnote{The lack of detections from transit surveys of clusters was not entirely expected \citep[see, e.g.][]{vansadersgaudi, masudawinn}.} but found success after the launch of the \Kepler\ space telescope, which detected two sub-Neptunes in the billion-year-old NCG 6811 cluster during its original mission \citep{meibom}. The turning point for detecting planets in clusters came when the failure of a second reaction wheel ended the original \Kepler\ mission and forced the spacecraft to point towards the ecliptic plane to maintain stable pointing \citep{howell}. Fortuitously, a wealth of nearby and well-studied clusters and associations, including the Hyades, Praesepe, Pleiades, M67, Ruprecht 147, and Upper Scorpius, happen to lie near the ecliptic plane, making \Kepler's extended K2 mission well suited for detecting small transiting planets around these well-characterized stars. K2 has fulfilled that promise with the detection of four planets in the Hyades \citep{mann16, davidhyades, mann2018,ciardi, livingston}, six planets and one candidate in Praesepe \citep{obermeier, libralato, Mann:2017aa}, one planet in Upper Scorpius  \citep{mann16b, david}, one planet in the Cas Tau association \citep{castau}, and one planet in Ruprecht 147 \citep{curtis}. 

The sample of small transiting planets in open clusters is already showing intriguing patterns, perhaps hinting that planets in young clusters may be less dense than their older counterparts \citep{mann16, obermeier, Mann:2017aa}. However, the  inferences which might be made about the existing population of planets in open clusters are limited by the sample. Because of its short observing baseline, K2 is most sensitive to planets in periods less than about 40 days, so the known small transiting cluster planets tend to orbit close to their host stars and be highly irradiated. Meanwhile, although radial velocity surveys have detected some long-period, cool planets, these objects are quite massive. Currently, there are no known small planets in temperate orbits around stars in open clusters, making it impossible to study the evolution and properties of planets in low-irradiation regimes. 


Here we report the detection of a long-period transiting planet candidate around the bright {\ron (V=10.6, K=7.7, Kp=10.1)} Hyades member \thisstar. We detected a single transit event in K2 Campaign 13 observations of \thisstar, with a depth, duration, and shape corresponding to a super-Earth in a roughly 100 day orbit around a K-dwarf stellar host. If the candidate is eventually confirmed to be real, it would be the first known temperate small planet in an open cluster. Our paper is organized as follows: in Section \ref{observations}, we describe the K2 discovery observations and both archival and follow-up data on \thisstar. Though we do not validate that the candid-eps-converted-to.pdfate is indeed an exoplanet with high confidence, our analysis of K2 data, spectroscopy, and imaging suggests this is likely the case. In Section \ref{analysis}, we perform an analysis to determine stellar and planetary parameters under the assumption that the single transit event we see is indeed due to an exoplanet. In Section \ref{discussion}, we discuss the uniqueness of the candidate around \thisstar\ and explore the path towards confirming the transits to enable further study.

\begin{figure*}[ht!] 
   \centering
   \includegraphics[width=6.5in]{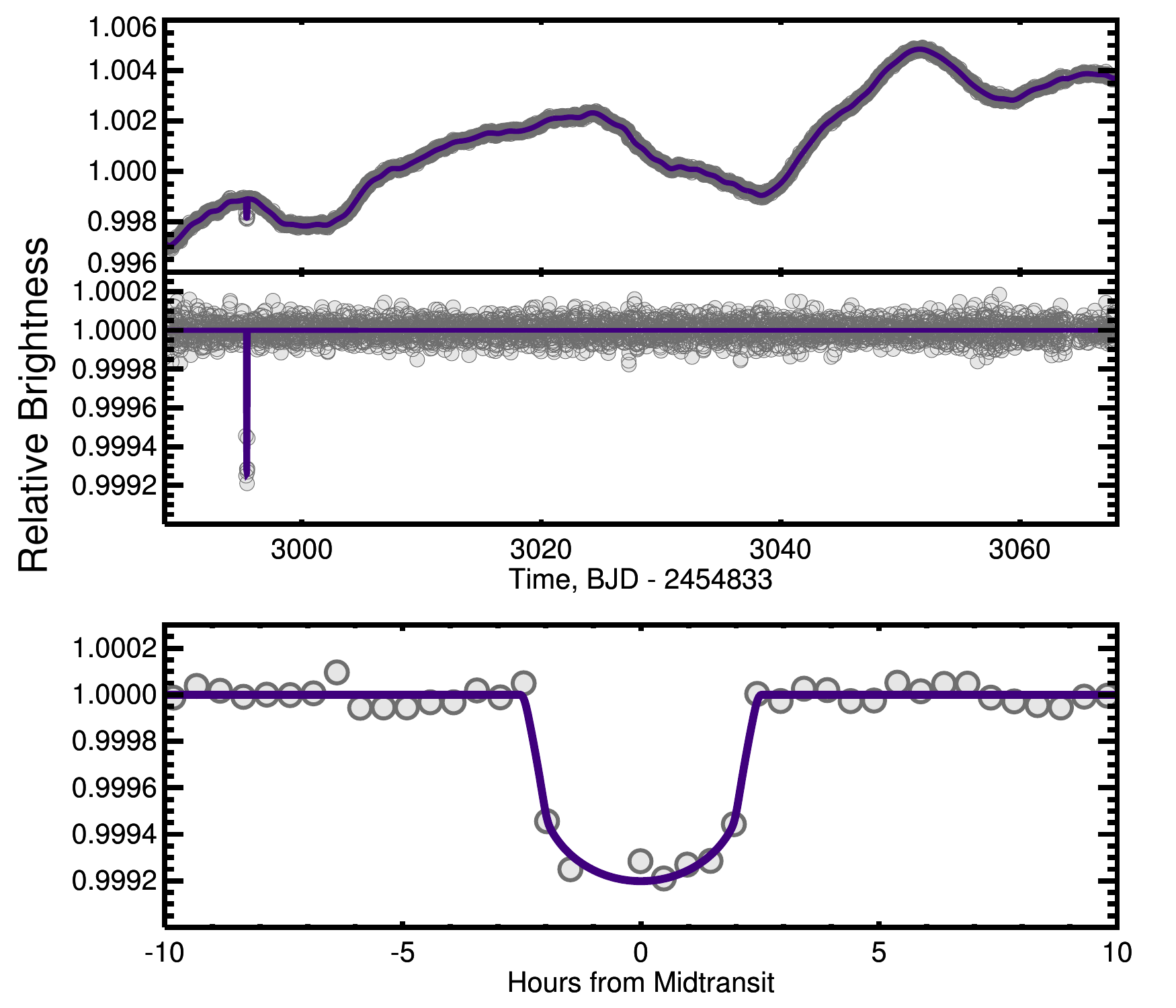} 
   \caption{Top: Systematics corrected K2 light curve of \thisstar. Grey circular points are the individual K2 long-cadence flux measurements and the purple curve is the best-fit low-frequency variability and transit model. The star shows variability with a period of about 37 days -- significantly longer than most other Hyades members of this mass -- and a single-transit event at time $BJD-2454833 \approx 2995$. Middle: K2 light curve with stellar variability removed. The transit signal is clearly visible, significant at the $\approx 20 \sigma$ level. Bottom: Zoom-in of the flattened K2 light curve with best-fit transit model overlaid. The signal is consistent with the transit of a super-Earth-sized exoplanet with an orbital period of about 100 days.}
   \label{lc}
\end{figure*}

\section{Observations}\label{observations}

\subsection{K2 Light Curve}\label{lightcurve}

\Kepler\ observed part of the Hyades cluster, including \thisstar, designated EPIC 248045685, during the 13th campaign of its extended K2 mission between 2017 March 8 and 2017 May 27. After the data were downlinked from the spacecraft, they {\ron were} processed by the K2 mission pipeline and released to the public. We downloaded the calibrated target pixel files from the Mikulski Archive for Space Telescopes, produced light curves, and removed systematic errors caused by \Kepler's unstable pointing using the method described by \citet{vj14}. We searched the processed light curves for transits using a Box-Least-Squares algorithm \citep{kovacs, v15}. Even though our transit search algorithm is designed to identify periodic phenomena, it detected a single, high signal-to-noise\footnote{We estimate the signal-to-noise of the dip is roughly 21.} transit-like dip in the brightness of \thisstar. The dip had a depth of about 800 ppm, a duration of about 4.6 hours, and a shape characterized by a rounded bottom and short ingress and egress times, consistent with the transit of a small exoplanet.  

Upon identifying the transit-like event, we re-processed the K2 light curve by fitting a systematics model simultaneously with the long-timescale variability of the star and a single transit of a long-period planet \citep[see][for details]{v15}. Our final K2 light curve is shown in Figure \ref{lc}. The K2 light curve is dominated by a long-period signal, which we think is likely astrophysical and could be related to stellar rotation. We measured a period of about $37 \pm 2$ days in the K2 light curve using both an autocorrelation function and Lomb-Scargle analysis. If this period is in fact the rotation period of the star, then \thisstar\ is an anomalously slow rotator for a star of its mass and age; most single Hyades and Praesepe members with similar masses have rotation periods of about 10-15 days. We discuss this point further in Section \ref{discussion}.  When the long-period signal is removed, the dip is clearly visible by eye near the beginning of the K2 light curve. 

\begin{figure*}[ht!] 
   \centering
   \includegraphics[width=6.5in]{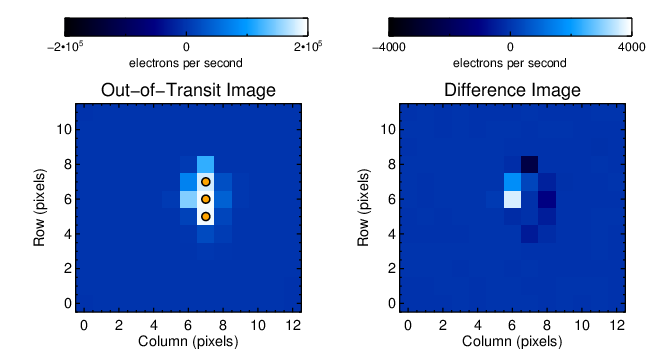} 
   \caption{{\ron Difference image analysis for the candidate transit event around \thisstar. Left: Out-of-transit image taken shortly after transit when \Kepler\ was at nearly the same position in its roll as during the middle of the candidate transit event. The orange dots mark three saturated pixels (with electron fluxes greater than about $1.6\times10^{5}$ per second, \citealt{keplerinstrumenthandbook}). Right: Difference image calculated by subtracting a K2 image taken during the transit from the out-of-transit image. While difference images for saturated stars observed by \Kepler\ are tricky to interpret, the source of the transit is on target. The morphology of the difference image is consistent with a genuine transit of \thisstar.}}
   \label{differenceimage}
\end{figure*}

While K2 data are typically quite reliable, occasionally single events like the one we detect in the light curve of \thisstar\ can be caused by instrumental phenomena. We therefore subjected the single dip to a battery of tests to rule out various scenarios which we have observed to cause similar signals in K2 data in the past. In particular, we confirmed that there were no changes to the scattered background light (perhaps caused by a bright Solar System object moving across \Kepler's focal plane\footnote{For an example of such a scenario see Figure 4b of \citet{v14}, which shows a spurious single transit-like event caused by an increase in scattered background light as the planet Jupiter moved out of \Kepler's focal plane.}) during the 4.6 hour transit-like event. We also confirmed that the dip was not a residual of our correction for systematics caused by K2's repeated drifting motion and thruster corrections. The dip spanned two drift periods and took place while \Kepler\ was oriented in a part of its roll that was well-characterized by our ``self flat field'' systematics correction. We also inspected the light curves of the two other stars\footnote{In particular, \url{https://archive.stsci.edu/prepds/k2sff/html/c13/ep248053336.html} and \url{https://archive.stsci.edu/prepds/k2sff/html/c13/ep248053424.html}.} observed by K2 within 5 arcminutes of \thisstar\ and found no similar simultaneous dips, indicating that the transit-like-event was not caused by some wide-reaching detector anomaly. {\ron We performed standard K2 pixel-level tests \citep[see, e.g.][]{v15, mayo} and confirmed that the apparent position of the star did not shift appreciably during the transit-like event both by difference image analysis (see Figure \ref{differenceimage}) and analysis of measured image centroids\footnote{{\ron With a \Kepler-band magnitude of 10.15, the image of \thisstar\ is saturated in the K2 images, which can confuse diagnostics like image centroid shifts and difference images. Nevertheless, with the difference image analysis, we are able to show that the source of the transit is cospatial with \thisstar, and we are able to confirm that the shift in image centroids (transverse to the spacecraft roll) during transit is less than about 2 milliarcseconds compared to the spacecraft position in the two days surrounding the transit. }}. Finally, we showed that the shape and depth of the transit remained the same when the photometric aperture used to extract the light curve was changed.}

Based on these tests, we conclude that the transit-like event we see is probably caused by some astrophysical phenomenon in the direction of \thisstar, and throughout the rest of the paper, we proceed under this assumption. In Sections \ref{spectroscopy} and \ref{imaging}, we go further and argue that that the most likely explanation for the dip in the light curve of \thisstar\ is that the star is indeed transited by a small, long-period exoplanet, but we do not go so far as to attempt to validate the signal as being caused by a genuine exoplanet with high confidence. Instead, given the difficulty of ruling out all possible false positive scenarios for single transit events, we consider the likely source of the signal to be a ``planet candidate,'' which it will remain until it is confirmed by the detection of additional transits or through precise Doppler monitoring \citep[e.g.][]{hip116454}. For convenience, throughout the rest of the paper, we refer to the planet candidate as \thisplanet.

\subsection{Spectroscopy}\label{spectroscopy}

\thisstar\ is a well studied star thanks to its long-suspected Hyades membership. Here, we make use of extensive archival observations and some new observations taken after we identified the planet candidate orbiting \thisstar. 

After being identified as a candidate Hyades member by photometric and proper motion surveys, \thisstar\ was observed spectroscopically three times between 1974 and 1980 with the Radial Velocity Spectrometer at the Coud\'e focus of the 5.1m Palomar Hale telescope \citep{griffin} as part of a survey to identify true Hyades members among previously identified candidates. The three RV measurements from this survey had a mean velocity of 39.6 $\pm$ 0.17 \kms on the IAU system\footnote{\citet{griffin} measured a mean velocity of 40.3 \kms. We offset the \citet{griffin} velocities to the IAU system by applying a correction of -0.84 \kms\ between the \citet{griffin} system and the CfA Digital Speedometer system, which we derived from observations of constant-velocity targets in common between the \citet{griffin} and CfA programs. Once the velocities were on the CfA system, we shifted them to the IAU system by applying a correction of +0.14 \kms.} (with no variations at the 500 \ms\ level), suggesting kinematics consistent with Hyades membership\footnote{The mean Hyades radial velocity is 39.3 $\pm$ 0.25 \kms\ with a velocity dispersion is 2.8 \kms\ \citep{Mermilliod}.}. 

Some of us began observing \thisstar\ in 1991 as part of an RV survey of Hyades members using the CfA Digital Speedometers on the 1.5m Wyeth Reflector at Oak Ridge Observatory in the town of Harvard, MA and on the 1.5m Tillinghast Reflector at Fred L. Whipple Observatory on Mt. Hopkins, AZ \citep{stefanik}. We obtained a total of 17 observations with the CfA Digital Speedometers between 1991 and 2006, all but two of which came from Oak Ridge Observatory. The RV time series shows no convincing evidence for astrophysical variability at the 300 \ms\ level, and a periodogram search reveals no strong periodicities. The mean velocity of the 17 Digital Speedometer observations is 39.7 $\pm$ 0.13 \kms\ on the IAU scale. There is no significant velocity difference between the CfA observations and the Palomar observations taken two decades earlier. 

More recently, we observed \thisstar\ with the Tillinghast Reflector Echelle Spectrograph (TRES), the high-resolution successor to the CfA Digital Speedometers on the 1.5m telescope at Mt. Hopkins. We obtained one observation in October 2011 and two other observations in September 2017 after we identified the planet candidate. We measured relative radial velocities between the three TRES observations using methods developed by \citet{buchhave10}. We detect a possible 80 \ms\ RV shift between the observation taken in 2011 and the two observations taken in 2017, but the formal confidence of this shift is only about 2$\sigma$, and we do not consider it significant. When placed on the IAU scale, the average of the three TRES RVs is 39.84 $\pm$ 0.1 \kms, where the uncertainty is dominated by the  transfer onto the IAU system. We adopt this value for the absolute RV. 

The most precise existing radial velocity observations of \thisstar\ were conducted as part of a survey to detect giant planets in the Hyades using the High Resolution Echelle Spectrograph (HIRES) on the 10m Keck I telescope on Maunakea, HI \citep{cochran, paulson}. \thisstar\ was observed six times between 1998 and 2003 with typical uncertainties of about 5 \ms. We placed limits on radial acceleration on \thisstar\ by fitting the six HIRES RV measurements with a linear model while allowing for a radial velocity ``jitter'' term. We found no statistically significant acceleration, measuring a best-fit slope of about 3 $\pm$ 2 \msy, roughly the acceleration induced by either a Jupiter mass planet at 5 AU, or an 0.1 \msun\ M-dwarf at 50 AU. Significantly closer or more-massive objects than this must be nearly face-on in order to escape detection.  

All in all, four decades of spectroscopic observations of \thisstar\ show no evidence for radial velocity variations, placing strong limits on the presence of binary companions. The lack of detected RV variations show definitely that \thisstar\ is not a short-period eclipsing binary, eliminating that false positive scenario for the planet candidate. The RV constraints also place limits on the presence of distant companions which might be eclipsing systems themselves, decreasing the likelihood of a hierarchical eclipsing binary false positive scenario. 

\subsection{Imaging}\label{imaging}

\begin{figure*}[ht!] 
   \centering
   \includegraphics[width=6.5in]{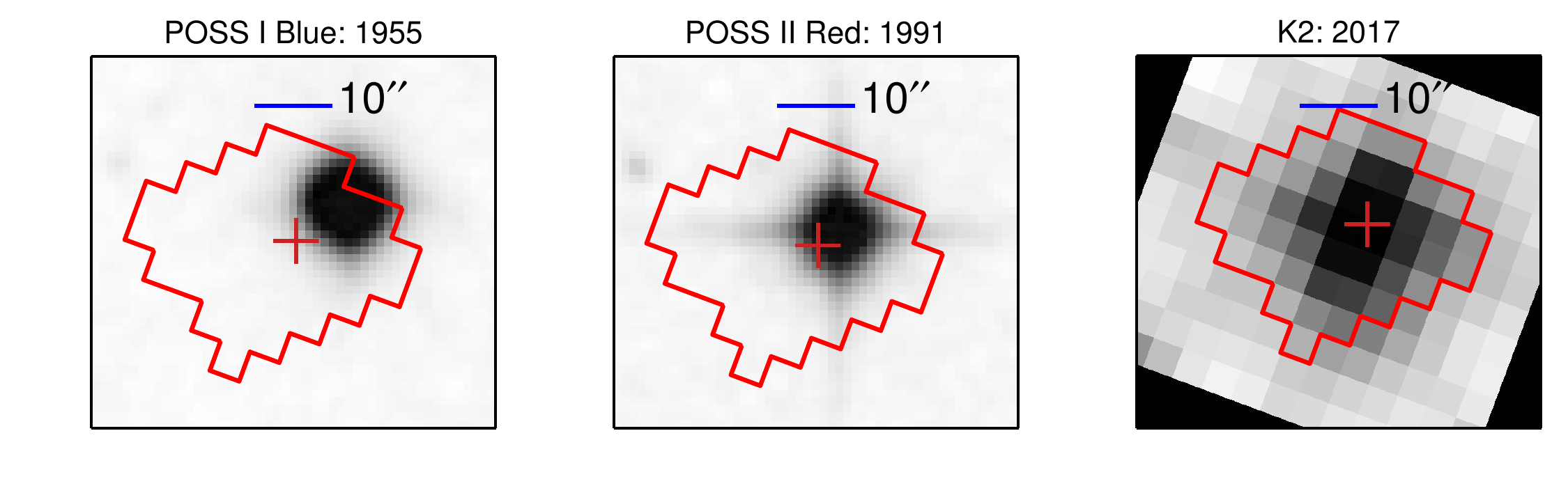} 
   \caption{{\ron Archival imaging of \thisstar. In these images, the outline of the K2 photometric aperture is shown as a red polygon and the present-day position of \thisstar\ is shown as a red cross near the center of the images. The 10\arcsec\ blue horizontal line near the top of the images shows the scale. Left: Image from the POSS I survey taken in 1955 on a photographic plate with a blue-sensitive emulsion. Middle: Image from the POSS II survey taken in 1991 on a photographic plate with a red-sensitive emulsion. Right: Summed image from the Campaign 13 K2 observations. The high proper motion of \thisstar\ makes it possible to exclude bright background companions at the star's present day location.}}
   \label{archivalimages}
\end{figure*}

We used a combination of archival imaging and newly acquired high angular resolution images to search for visual companions to \thisstar. We first inspected images taken in the original Palomar Observatory Sky Survey (POSS) on a photographic plate with a blue-sensitive emulsion to search for stationary background objects close to the present day position of \thisstar. Since \thisstar\ was observed by POSS in 1955, its apparent position in the sky has moved by about 9 arcseconds, making it possible to search for stationary background stars near the its present-day position {\ron(see Figure \ref{archivalimages})}. In a blue-sensitive plate, the saturated point spread function of \thisstar\ extends near its present-day position 9 arcseconds away, and we see no evidence for any elongation that might hint at a background star in the present-day location of \thisstar. We estimate based on the other nearby faint stars in the POSS image that if there was a star brighter than about 18th magnitude at the present-day position of \thisstar, we would have seen it. Since we see no such star close to the present-day position of \thisstar, we can exclude background stars about 6 magnitudes fainter in blue bandpasses. We also searched for wide co-moving binary companions using the Hot Stuff for One Year (HSOY) catalog \citep{hsoy}. We identified no other stars out to a distance of 900 arcseconds (about 40,000 AU projected distance) brighter than R$\approx$19 (corresponding to roughly 0.1 \msun\ M-dwarfs) with a proper motion consistent with \thisstar. {\ron Finally, we queried the \Gaia\ DR2 database for sources near \thisstar\ \citep{gaiamission, gaiadr2}. \Gaia\ identified three very faint point sources within the K2 photometric aperture at distance of 3.7\arcsec, 9.2\arcsec, and 12.8\arcsec. These point sources are too faint for \Gaia\ to have measured proper motions or parallaxes, so we cannot ascertain whether any of them are physically associated with \thisstar\ or if they are background objects. All three of these stars have \Gaia-band G magnitudes fainter than G=19.4, too faint to have caused the 700 ppm transit signal we observed on \thisstar. Evidently, there are no widely separated stars near \thisstar\ which could have contributed the transit signal we see.}

After identifying the planet candidate, we observed \thisstar\ with two speckle imaging instruments: the NN-Explore Exoplanet Stellar Speckle Imager (NESSI) on the 3.5m WIYN telescope on Kitt Peak in Arizona, and \Alopeke\ on the 8m Gemini-N telescope on Maunakea, HI. NESSI and \Alopeke\ both work by taking many short (40-60 ms) exposures of a target star simultaneously in two optical narrow bands. The short exposures freeze out atmospheric turbulence, resulting in sub-images which can be reconstructed using Fourier techniques to produce diffraction-limited images over small fields of view. We observed with NESSI in 40 nm-wide filters centered at 562 and 832 nm and with \Alopeke\ in similar filters centered at 562 and 880 nm\footnote{Due to poor weather conditions for our observation with \Alopeke, only the image taken with the 880 nm filter was usable.}.We reduced the data using the method described by \citet{howell2011}, and detected no nearby companions in any of the reconstructed images. The strongest constraints at small angular separations are placed by \Alopeke; we can exclude stars 4.4 magnitudes fainter at angular separations of 0.1 arcseconds (or projected distances of 5 AU). The NESSI images are deeper than the \Alopeke\ images due to observing conditions, and contribute the strongest constraints at larger angular distances. The NESSI data at 832 nm exclude stars about 5.8 magnitudes fainter at this wavelength at distances of about 1 arcsecond, or projected distances of 50 AU. 

The constraints we place on background objects and visual companions from archival and speckle imaging further limit false positive scenarios, making it more likely that the planet candidate around \thisstar\ is indeed a transiting exoplanet. Therefore, throughout the rest of this paper, we perform analyses assuming that \thisstar\ is single and that the candidate transit event is indeed caused by a transiting exoplanet.

\section{Analysis}
\label{analysis}
\subsection{Membership in the Hyades}

\thisstar\ has a long history of being associated with the Hyades cluster. \citet{griffin} measured a radial velocity for \thisstar\ consistent with Hyades membership, but they flagged it as a possible member, citing inconsistencies in literature proper motion measurements as a source of doubt. More recently, \citet{perryman}  and \citet{roser} assigned \thisstar\ membership using updated astrometric parameters from \hipparcos\ \citep{esahipparcos} and the PPMXL catalogs, respectively.  

\begin{figure*}[t!] 
   \centering
   \includegraphics[width=6.5in]{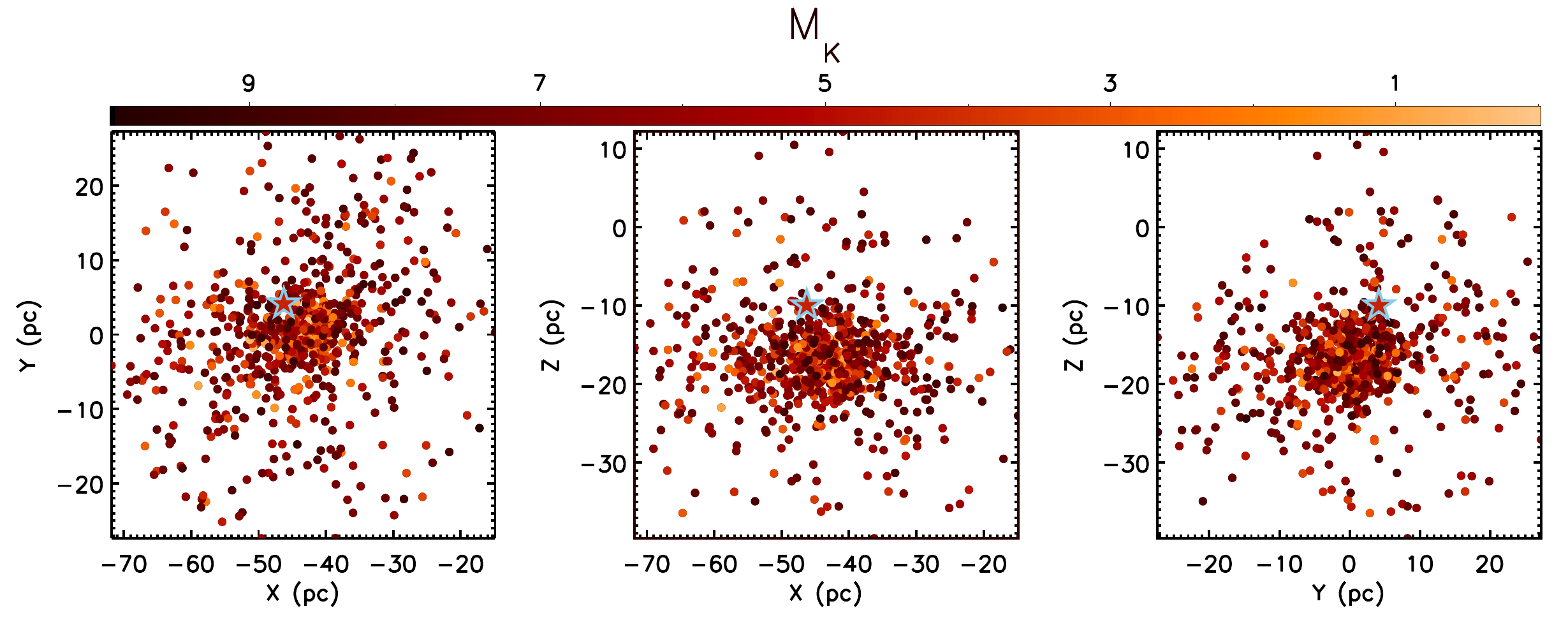} 
   \caption{Galactic coordinates of \thisstar\ in relation to other known Hyades members. The colors of the points correspond to the absolute  \thisstar\ is shown as a grey star, while the other members are shown as colored circular points, with the color corresponding to each star's absolute K-band magnitude. \thisstar's position is near the edge of the Hyades core, well within the larger distribution of Hyades members.}
   \label{position}
\end{figure*}

We reassessed the case for \thisstar's membership in the Hyades. First, we note that there is solid evidence for \thisstar's membership based on its position and proper motion. \thisstar\ is located near the outskirts of the Hyades core (see Figure \ref{position}), and the star's space velocity is towards the cluster's convergence point \citep[The star has a velocity of 23.7 \kms\ parallel to the cluster's convergence point and only 1.3 \kms\ perpendicular to the convergence point, ][]{roser}. Using the methods described by \citet{rizzuto11} and \citet{rizzuto15}, and the Hyades cluster model from \citet{rizzuto}, we calculate a membership probability greater than 99\%. This calculation does not take into account the measured radial velocity (consistent with Hyades membership) and the fact that \thisstar\ falls right on the Hyades main sequence in a color-magnitude diagram. Including this additional information brings the membership probability to near unity. Although \thisstar\ has a slightly discrepant proper motion perpindicular to the cluster convergence point (larger than all but a handful of other known members) and might have an anomalously long rotation period (see Section \ref{slowrotation}), the preponderance of the evidence suggests that it is indeed a Hyades member. 

\subsection{Limits on Additional Transiting Planets} \label{limits}

\begin{figure}[t!] 
   \centering
   \includegraphics[width=3.5in]{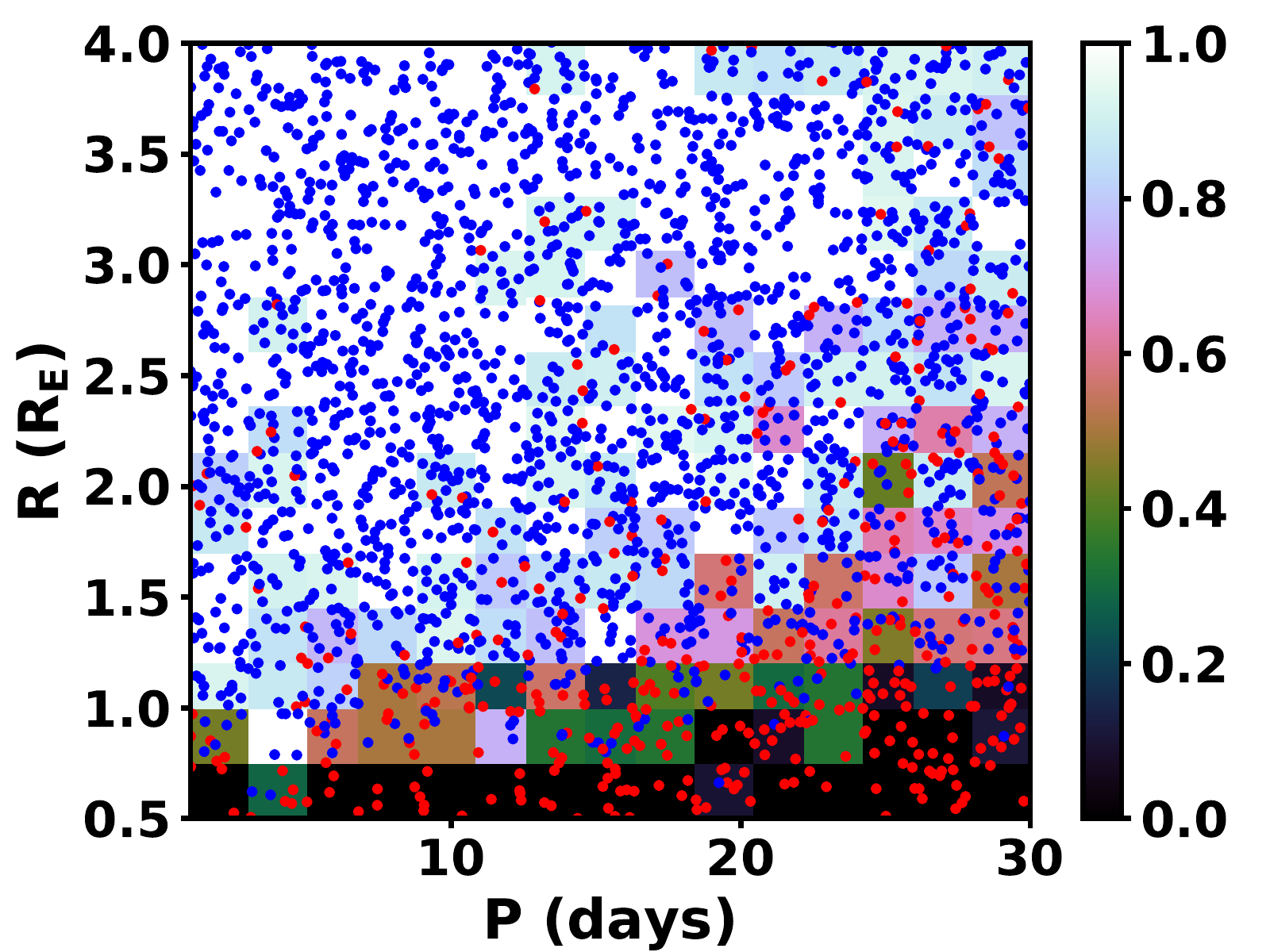} 
   \caption{Sensitivity to additional transiting planets around \thisstar. We show the orbital periods and planet radii of our injected planets as circular points in the plot; blue points represent planets which we successfully recovered with our notch-filter pipeline, and red points indicate planets which we did not recover. The plot background color shows the fraction of recovered planets in each region of parameter space.}
   \label{injected}
\end{figure}

We placed limits on additional (short-period) transiting planets by performing injection/recovery tests following the procedure outlined by \citet{rizzuto}. We injected 4000 transit signals with randomly chosen planet and orbital parameters into the light curve of \thisstar\ and attempted to recover them with the ``notch-filter'' pipeline described by \citet{rizzuto}. Our results are shown in Figure \ref{injected}. We find that we are generally sensitive to sub-Earth-sized planets in short-period ($\lesssim 5$ day) orbits and somewhat sensitive to Earth-sized planets out to periods of about 25 days.  If there are other similarly-sized planets orbiting interior to \thisplanet, then there must be some misalignment between the planets' orbits. 

\subsection{Stellar Parameters}\label{stellarparameters}

We used the Stellar Parameter Classification \citep[SPC,][]{buchhave, buchhave14} method to determine the effective temperature, surface gravity, and equatorial rotational velocity of \thisstar\ from the three TRES spectra. We ran SPC while fixing the metallicity to the cluster metallicity; we used a value of +0.15 which is an average of several previous determinations \citep{paulson2003,dutraferreira}. Averaging the results for each of the three spectra, we measure a temperature $T_{\rm eff,SPC} = 4686 \pm 50$ K, surface gravity $\log{g_{\rm SPC}} = 4.70 \pm 0.1$, and we place an upper limit on the star's projected equatorial rotation velocity of about 2 \kms. {\ron We measure an average Mt. Wilson activity $R'_{\rm HK}$ indicator from our three TRES spectra of $R'_{\rm HK} = -4.77 \pm 0.05$ using the procedure described by \citet{mayo}.} 

We estimated the luminosity of \thisstar\ using the parallax from \Gaia\ DR1 \citep[21.05$\pm$0.29\,mas,][]{gaiadr1}\footnote{{\ron Recently, a more precise parallax for \thisstar\ was included in \Gaia\ DR2 of 21.003$\pm$0.054\,mas. We confirmed that the stellar parameters and uncertainties derived using this new parallax remain consistent within errors, and the uncertainties in stellar parameters, which are dominated by systematic errors in stellar evolutionary models, were unchanged.}} and fitting empirical templates to the available photometry, following the procedure from \citet{Mann2015b} and \citet{Mann:2017aa}, which we briefly describe here. We first downloaded archive photometry from the literature, including $J\,H\,K_S$ from the Two Micron All Sky Survey \citep[2MASS,][]{Skrutskie2006}, $B_T$ and $V_T$ from Tycho-2 \citep{Hog2000}, $H_P$ from Hipparcos \citep{1997A&A...323L..61V}, $U\,B\,V$ from the General Catalogue of Photometric Data \citep[GCPD,][]{Mermilliod1997}, $B\,V$ and $r'$ from the AAVSO All-Sky Photometric Survey \citep[APASS,][]{Henden:2012fk}, $r'$ from the Carlsberg Meridian Catalogue \citep[CMC15,][]{CMC15}, and $W1\,W2\,W3\,W4$ from the Wide-field Infrared Survey Explorer \citep[WISE, ][]{Wright2010}. 

\begin{figure*} 
   \centering
   \includegraphics[width=3.5in]{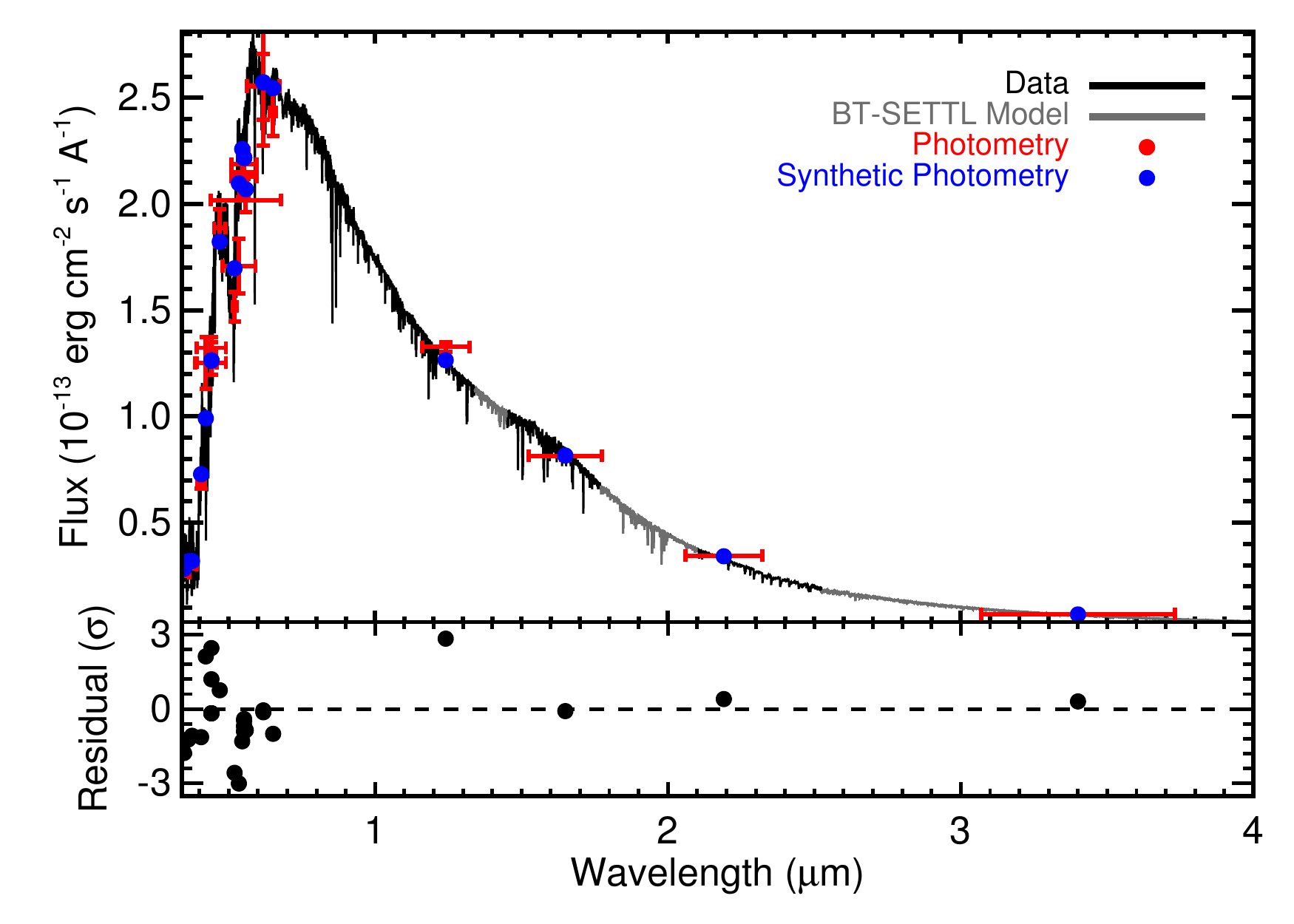} 
   \includegraphics[width=3.5in]{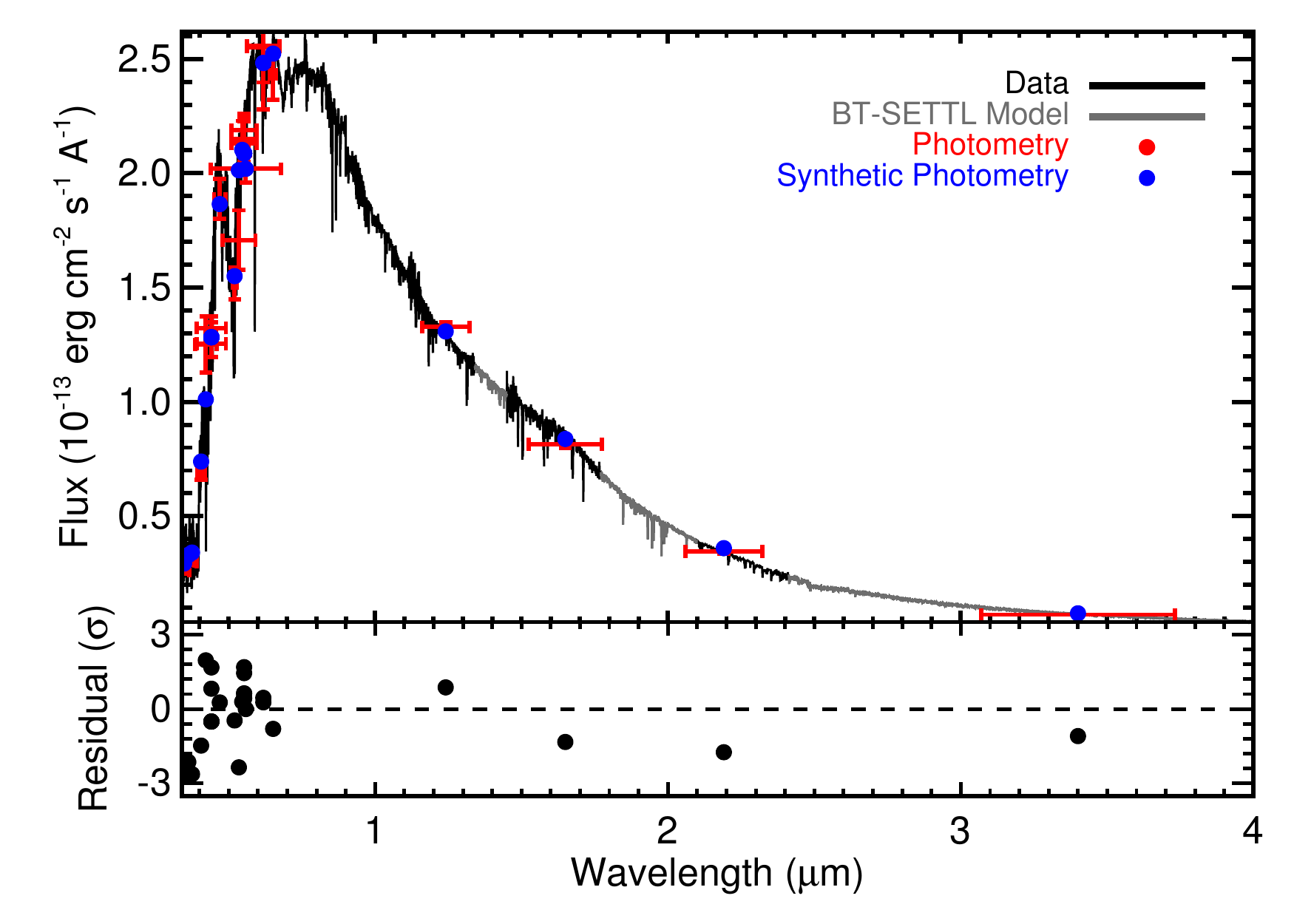} 
   \caption{Spectral energy distribution of \thisstar\ as a function of wavelength with the two best-fit templates (black, K5 on the left, and K7 on the right). Grey regions represent BT-SETTL models, which are used to fill in gaps in the templates. Literature photometry is shown in red, with vertical errors representing measurement uncertainties and horizontal errors an approximation of the filter width. Blue points represent synthetic photometry derived from the template spectrum. The bottom panel shows the residuals between observed and synthetic photometry in units of standard deviations. The K5 gives a slightly better fit (reduced $\chi^2$ of 1.7 and 2.0).}
   \label{fig:sed}
\end{figure*}

We converted literature photometry to fluxes using the appropriate filter profile and zero-point \citep[e.g.,][]{2003AJ....126.1090C,2012PASP..124..140B,Mann2015a}. Utilizing spectra from the IRTF Cool Stars Library \citep{Cushing:2005lr,Rayner2009} and CONCH-SHELL catalog \citep{Gaidos2014}, we found the best-fit spectral template by comparing these fluxes to values derived from these spectra, allowing the mean flux level of the template to float (Figure~\ref{fig:sed}). We filled in regions of high telluric contamination and those not covered by our templates using BT-SETTL models \citep{Allard2011}. Given that the star is within the `Local Bubble', reddening is likely to be negligible \citep{Lallement_Bubble}, and was not included in our analysis. The final bolometric flux was taken to be the integral over all wavelengths of the best-fit template and model, scaled to match the photometry. Interpolating between templates gave a negligible improvement in the fit (improvement in reduced $\chi^2$ of $<$0.1). Uncertainty on the bolometric flux was calculated by accounting for errors in the individual magnitudes, zero-points, and differences between templates. This procedure yielded a bolometric flux of $2.61\pm0.05 \times 10^{-9}$ erg\,cm$^{-2}$\,s$^{-1}$. Combined with the \Gaia\ DR1 parallax (21.05$\pm$0.29\,mas) this gave a luminosity of 0.182$\pm$0.006\,$L_{\odot}$.

To determine other stellar parameters, we interpolated this luminosity onto the Mesa Isochrones and Stellar Tracks \citep[MIST,][]{MIST0,MIST1} and Dartmouth Stellar Evolution Program \citep[DSEP,][]{Dotter2008}, using the canonical Hyades age \citep[600-800\,Myr][]{perryman,Brandt2015,2018arXiv180207155M} and metallicity \citep[$\simeq$0.15,][]{Liu2016}. Accounting for differences between the two model grids, and errors on the input parameters, this procedure gives $T_{\rm{eff}}$ = 4655$\pm$55\,K, $R_*=0.664\pm0.023M_\odot$, and $M_*=0.742\pm0.023M_\odot$. This $T_{\rm{eff}}$ is consistent with the value derived from the TRES spectrum. We also obtained a consistent radius using the Stefan-Boltzmann relation with the TRES $T_{\rm{eff}}$ and above luminosity, and a consistent mass using the empirical mass-luminosity relation from \citet{Henry:1993fk}, suggesting that the model-derived parameters are reasonable for this star. 

\subsection{Transit Light Curve}\label{transitanalysis}

We determined transit parameters by fitting the K2 light curve with a \citet{mandelagol} model\footnote{We accounted for the 29.4 minute \Kepler\ long-cadence integration time by oversampling the model light curve by factor of 30 and performing a trapezoidal integration.} using a Markov Chain Monte Carlo (MCMC) algorithm with affine invariant ensemble sampling \citep{goodman}. Often, when astronomers fit transits, they parameterize planetary orbits with physical variables such as the orbital inclination $i$ or the ratio of the planet's semi-major axis to the stellar radius $a/R_\star$. The large uncertainties and covariances in the orbital elements of singly-transiting planets make it difficult for MCMC explorations to converge in situations like that of \thisstar. Therefore, instead of using a physical parameterization, we fit the K2 light curve in terms of variables directly related to the shape of the transit. In particular, we fit the transit in terms of the planet-star radius ratio, $R_p/R_\star$, the full duration of the transit from first to fourth contact, $t_{14}$, the time of transit center $t_t$, the transit impact parameter, $b$, and linear and quadratic limb-darkening coefficients, $u_1$ and $u_2$. We also fit for a ``jitter'' term describing the uncertainty in the flux in each K2 long-cadence datapoint.  We imposed priors requiring both the transit duration and the flux uncertainty term to be greater than zero and requiring the impact parameter to be between 0 and 1 + $R_p/R_\star$. We imposed informative Gaussian priors on $u_1$ and $u_2$, centered on the values interpolated from limb darkening models \citep[0.644 and 0.096 for $u_1$ and $u_2$, respectively,][]{claretbloemen} with widths of 0.07, \citep[roughly matching the level of agreement between models and observations,][]{muller}. We explored parameter space with 100 walkers, which we evolved for 10,000 steps each, discarding the first half for burn-in. 

\subsection{Orbital Period}\label{orbperiod}

Because we only observed a single transit of the planet candidate \thisplanet, the candidate's orbital period is not well determined. We therefore estimated the orbital period of \thisplanet\ using a simplified version of the method described by \citet{hip41378}. We began by taking the posterior samples from our MCMC analysis of the K2 light curve described in Section \ref{transitanalysis}, which include 500,000 individual samples of the parameters \{$R_p/R_\star$, $t_{14}$, $b$\}. To estimate the orbital period of the planet, we took each set of these parameters drawn from the posterior, randomly drew samples of the eccentricity $e$ and argument of periastron $\omega$ from the joint distribution described by \citet{kipping_prior1} and \citet{kipping_prior2}, and calculated the orbital period $P$ by evaluating the following equation\footnote{This equation can be derived by simplifying Equation 2 from \citet{hip41378} if the scaled semimajor axis $a/R_\star \gg 1$, a safe assumption for long-period transiting planet candidates like \thisplanet.}: 

\begin{equation}
P = \left[ \frac{t_{14} (G M_\star \pi /4)^{1/3}}{\sqrt{(R_p + R_\star)^2 - b^2 R_\star^2}} \frac{1 + e \cos(\omega)}{\sqrt{1-e^2}} \right]^3
\end{equation}

\noindent where $G$ is the gravitational constant, $M_\star$ is the stellar mass, $R_p$ is the planetary radius, and $R_\star$ is the stellar radius. The resulting distribution of possible orbital periods for \thisplanet\ peaks at about 40 days, with long tails extending to short periods inside of 10 days and long periods well beyond one year. 

\begin{figure}[t!] 
   \centering
   \includegraphics[width=3.5in]{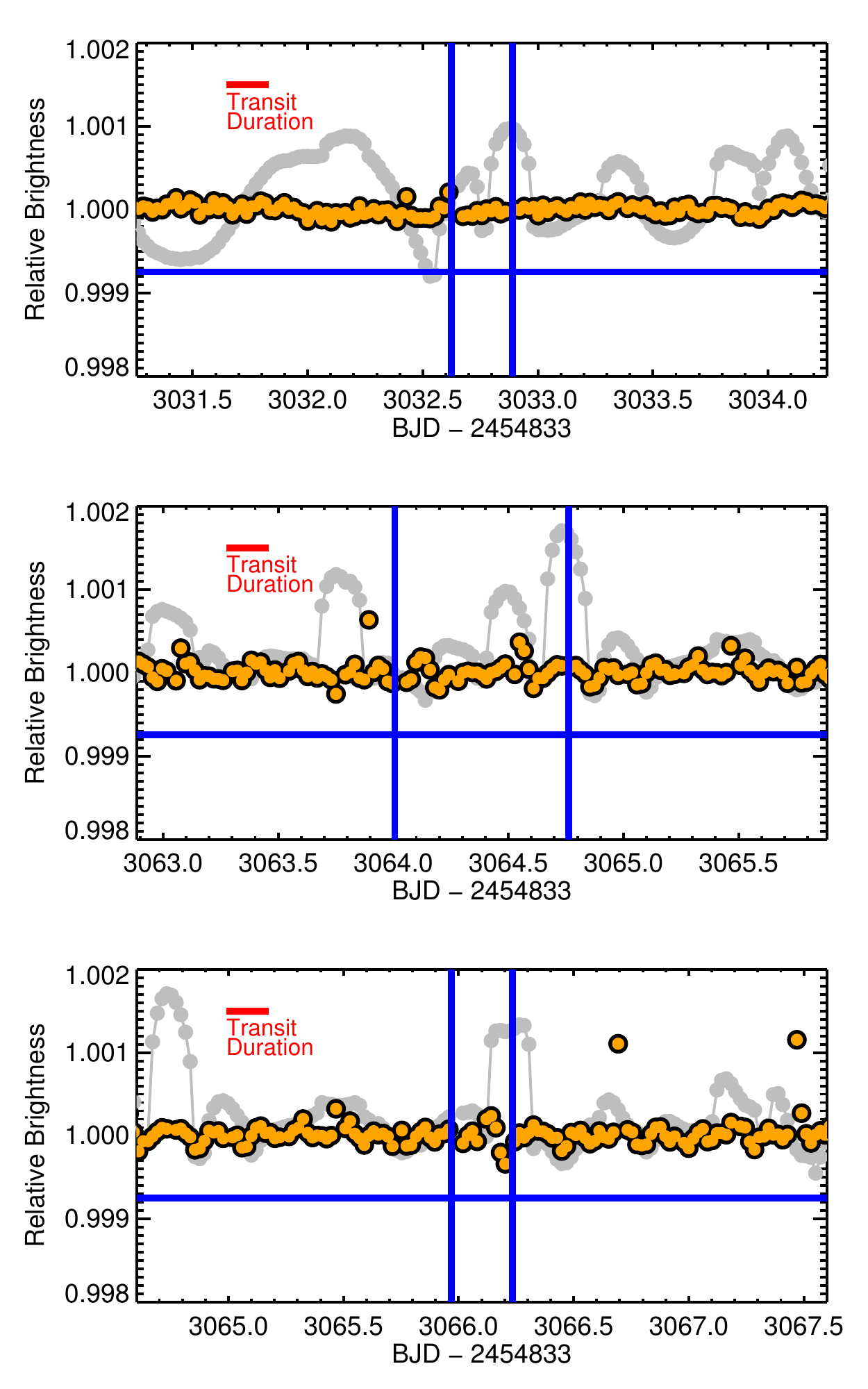} 
   \caption{{\ron K2 light curve during periods when the spacecraft lost fine-pointing control. Each panel shows both the systematics-corrected K2 light curve (orange) and the raw K2 light curve convolved with the shape of \thisplanet's transit (grey) to partially average over the uncorrected K2 roll systematics. We show the raw K2 light curve in addition to the more precise systematics-corrected light curve to demonstrate that no plausible transit signals were absorbed by the systematics correction in these poorly-constrained parts of the flat field. The periods when K2 lost fine-pointing control are interior to the two horizontal blue lines, and the depth of \thisplanet's transit is shown with the horizontal blue line. The duration of \thisplanet's transit is shown as a red horizontal line in the upper left-hand corner of each panel. There are no signals in either the raw or  systematics-corrected light curves during the periods without fine-pointing control consistent with a second transit of \thisplanet.}}
   \label{gaps}
\end{figure}

The duration, impact parameter, and planet-star radius ratio are not the only information we have at our disposal about the orbital period of \thisplanet. We can also place constraints based on the fact that the planet candidate only transited once during the 80 days of K2 observations. In particular, because the single transit occurred just about 8 days after the beginning of the K2 observations, and no other similar dips occurred during the rest of the observing campaign\footnote{While \Kepler\ observations during Campaign 13 were uninterrupted, our default light curve reduction excluded data from several short periods of time when the spacecraft briefly lost fine-pointing control. We re-reduced the K2 light curve while including these data and confirmed that no transits occurred during these gaps {\ron (see Figure \ref{gaps})}.}, the candidate's orbital period must be longer than about 72 days. We accounted for this by discarding all samples of the transit parameters and orbital periods with periods less than this minimum allowed period.

We also took into account the probability that we would detect the transit of a long-period planet at all in our observations. When the orbital period of a planet is longer than the duration of observations, there is no guarantee that the transit will take place while observations are taking place. For orbital periods longer than the duration of observations $B$, the probability $\mathcal{P}$ of detecting a transit decreases as:
\begin{equation}\label{priordetection}
\mathcal{P} = (B + t_{14})/P \text{  for  } P > B +  t_{14}
\end{equation}

We took this additional prior into account by randomly selecting whether to discard individual samples for periods longer than the observing baseline with a probability described by Equation \ref{priordetection}. 

\begin{figure}[t!] 
   \centering
   \includegraphics[width=3.5in]{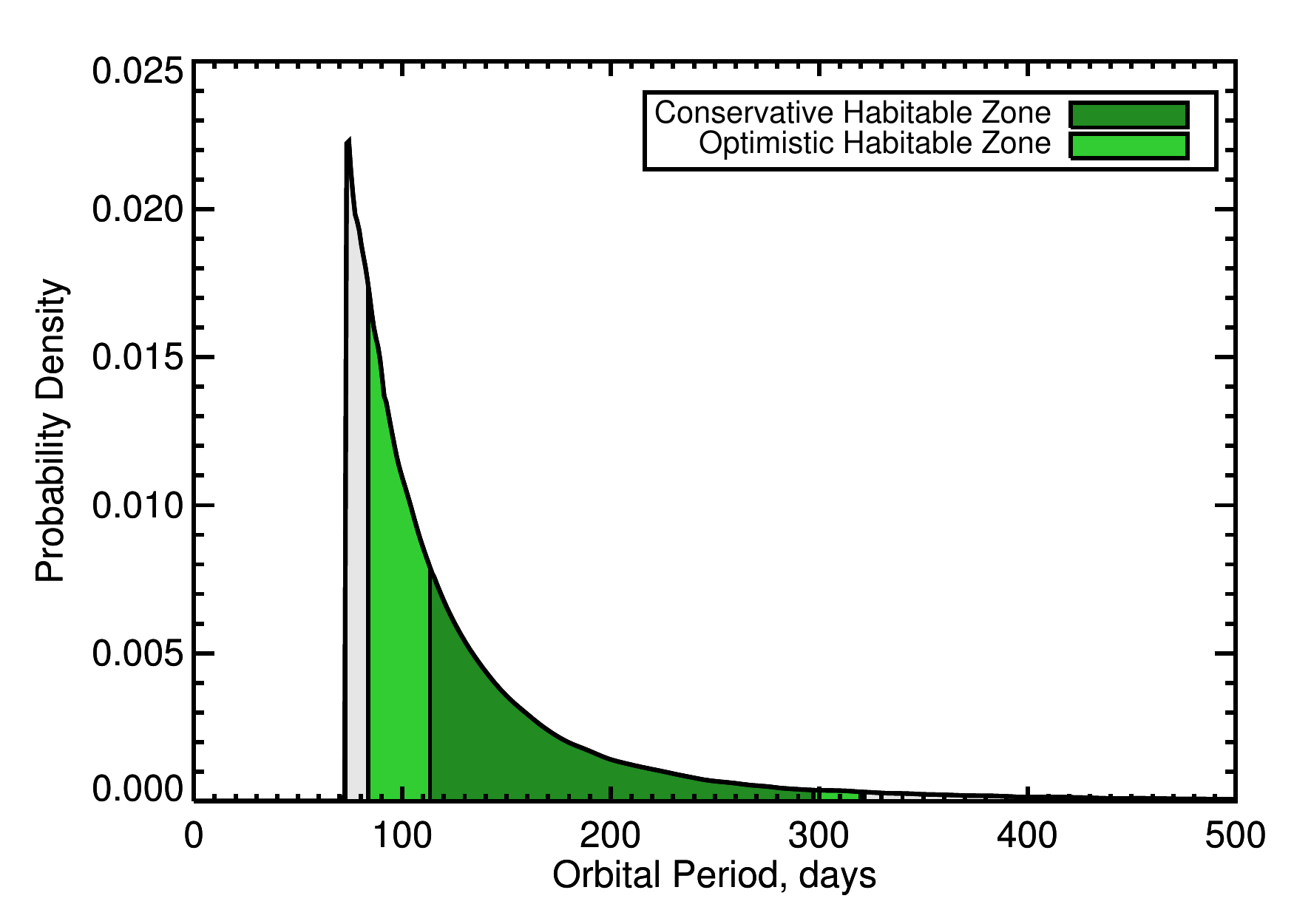} 
   \caption{Constraints on the orbital period of \thisplanet. The black curve shows the probability distribution of the orbital period from our analysis in Section \ref{orbperiod}. The light green and dark green shaded regions represent orbits which fall in the optimistic and conservative circumstellar habitable zones, respectively \citep{kopparapu}. Despite our weak constraint on orbital period, we can say fairly confidently that if real, \thisplanet\ is temperate, with a 71\% chance of orbiting within the star's habitable zone and a 99\% upper limit on equilibrium temperature of 327 Kelvin.}
   \label{period}
\end{figure}

We use the surviving samples to estimate both orbital and transit parameters for \thisplanet. The parameters are summarized in Table \ref{bigtable} and the orbital period probability distribution is shown in Figure \ref{period}. Most likely, the orbital period is not much longer than the minimum allowed period of 72 days; our analysis yields $P=\perplb\ \uperplb$ days\footnote{The orbital period is not particularly sensitive to the choice of eccentricity prior. If we assume the planet's orbit is circular, we find $P = 99^{+50}_{-20}$ days.}. Interestingly, given the luminosity and temperature of \thisstar, there is a fairly high likelihood that \thisplanet\ orbits in the host star's habitable zone. 71\% of the surviving orbital period samples fall within the optimistic habitable zone as calculated by \citet{kopparapu}, and 36\% of the surviving samples fall within the conservative habitable zone. The equilibrium temperature of \thisplanet\ is about $\teqb \uteqb$ Kelvin, which would make it the first temperate planet found in an open cluster. 

\section{Discussion}\label{discussion}

\subsection{Uniqueness of \thisplanet}
If confirmed to be real, \thisplanet\ would stand out among transiting planets in open clusters. With a K-band magnitude K=7.7, \thisstar\ would be the brightest star to host a transiting planet in a cluster, making detailed further studies possible. The brightness and slow rotation of \thisstar\ make it well suited for precise RV observations (though a detection of \thisplanet\ may have to wait for advances in the treatment of stellar activity, see Section \ref{massmeasurement}), and the brightness in the infrared and the fairly small size of the host star could make future transit transmission spectroscopy observations possible. 

What sets \thisplanet\ apart from the population of transiting planets in clusters is its long orbital period and low irradiation environment. The longest-period validated transiting planet in a cluster is K2-136 d \citep{mann2018}, which with a period of 25.6 days is the outermost planet in a three-planet system. \thisplanet\ likely has an orbital period more than three times longer than K2-136 d. \thisplanet\ would also be the transiting cluster planet which receives the least stellar irradiation. \thisplanet\ receives $1.2^{+0.5}_{-0.6}$ times the flux received by the Earth, four times less flux than is received by K2-103, the present record holder. 

The combination of its young age, proximity, and low-irradiation make \thisplanet\ an intriguing target for studying the development of small, temperate planets. At an age of roughly 600-800 million years, \thisplanet\ may still be evolving into its mature state. Radius evolution models calculated by \citet{lopezfortney14} for super-Earths with hydrogen-rich envelopes predict that in the absence of photoevaporation, if \thisplanet\ has a hydrogen-rich envelope, its radius will contract somewhere between 5\% and 10\% between now and maturity at an age of about 5 Gyr. Comparisons of the density of \thisplanet\ to similar planets around older field stars could test these models. Observations of \thisplanet\ might otherwise reveal surprises; other transiting planets discovered in the Hyades and Praesepe like K2-25 b and K2-95 b seem to be larger than their counterparts around mature stars \citep{mann16, obermeier, Mann:2017aa}, indicating that processes like atmospheric evaporation may still be taking place. If transit observations of \thisplanet\ show evidence for atmospheric loss, \thisplanet\ might be the progenitor of an even smaller temperate planet, and potentially an early version of a rocky habitable-zone planet.

\subsection{Evidently Slow Rotation}\label{slowrotation}

In Section \ref{lightcurve}, we identified a possible 37-day rotation period for \thisstar, which is considerably longer than the rotation periods of stars of similar mass and age in the Hyades and the similarly aged Praesepe open cluster. At face value, this is surprising. Several groups \citep{douglas16, douglas17, rebull17} have used K2 data to measure rotation periods of Hyades and Praesepe stars and found tight period-mass relations for single stars in these clusters, with high ($\approx85$\%) recovery fractions. A few other Hyades-age stars show longer-period variability than their peers, including the Praesepe member EPIC 211974724 with a 35 day period \citep{agueros, douglas17}, but it is unclear whether these long rotation periods are actually reliable. {\ron \thisstar\ also appears unusually inactive in spectroscopic indicators. For \thisstar, Mt. Wilson $R'_{\rm HK} = -4.77$, while the median $R'_{\rm HK}$ for Hyades stars is -4.47 with a scatter of 0.09 \citep{mamajek}. While \thisstar's H-$\alpha$ equivalent width is not easily distinguished from other Hyades age stars in low-resolution spectra obtained by \citet{douglas14}, inspection of high-resolution spectra of some of these stars shows \thisstar\ is less active in H-$\alpha$ as well.} 

\begin{figure*} 
   \centering
   \includegraphics[width=3.5in]{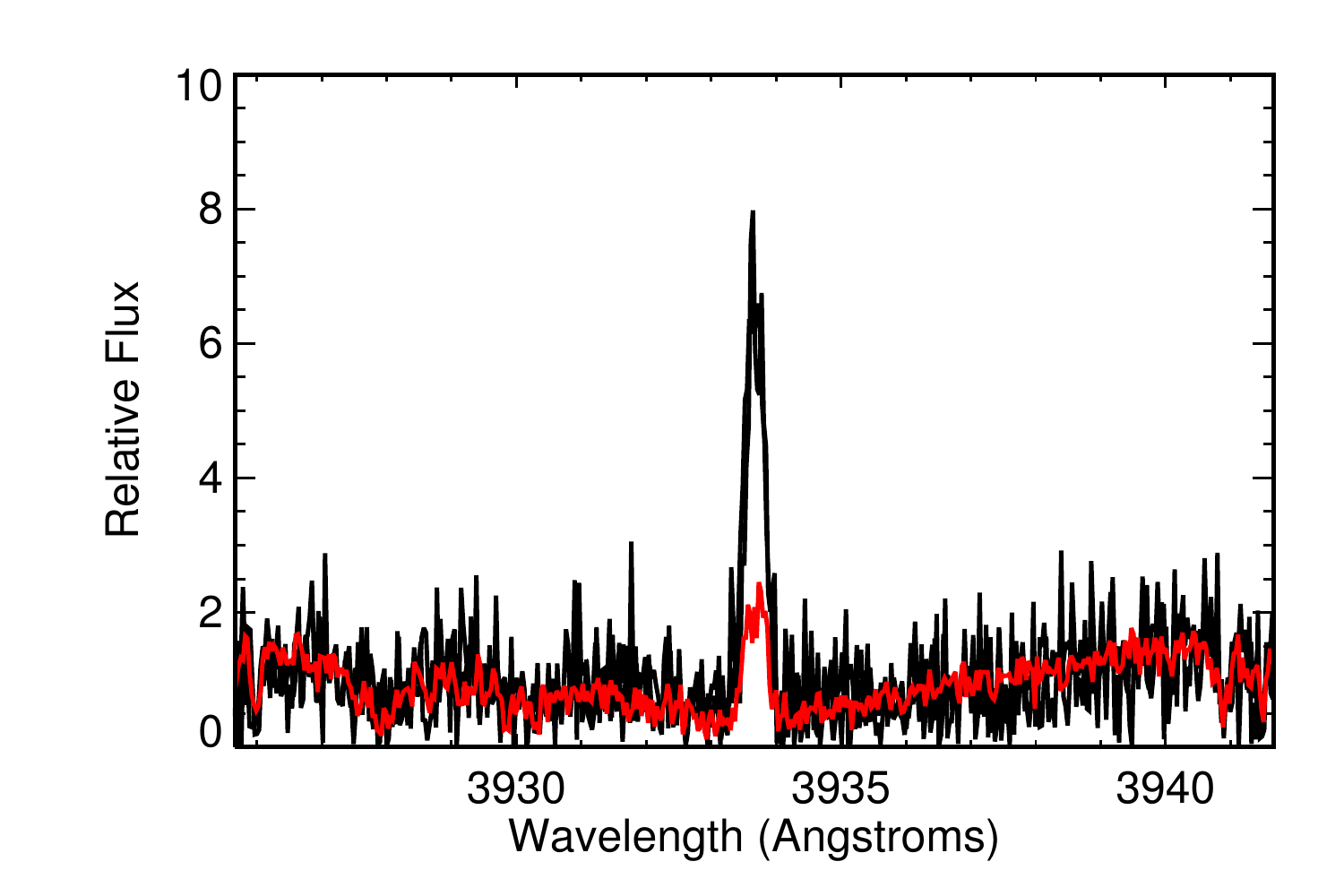} 
   \includegraphics[width=3.5in]{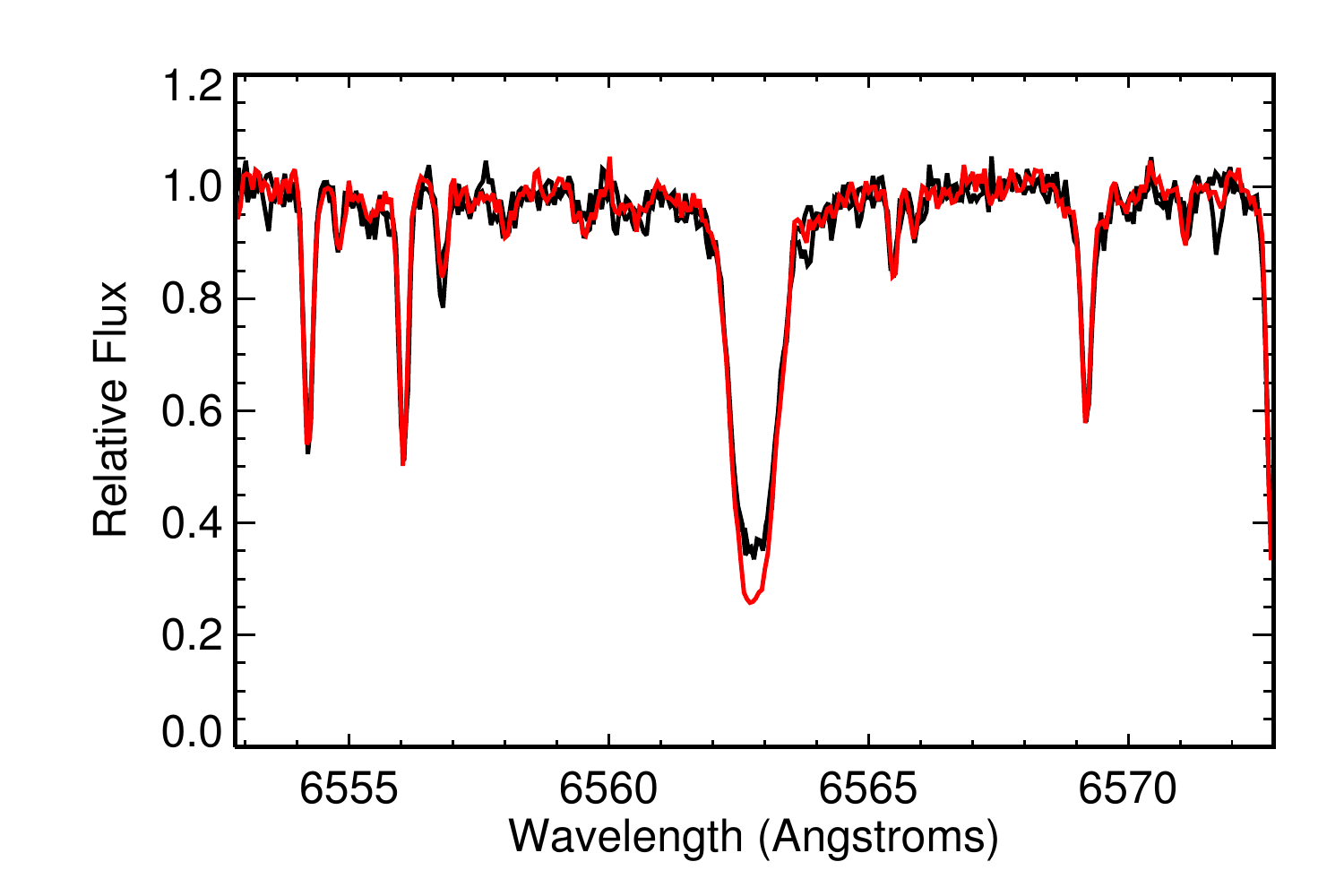} 
   \caption{{\ron Activity indicators of \thisstar\ compared to other Hyades stars of similar mass from TRES spectra. The left plot shows the Ca II K line in the ultraviolet, and the right plot shows H-$\alpha$. In both plots, spectra of \thisstar\ are shown in red, while the spectra of the other Hyades stars (HD 286572 and HD 286789) are shown in black. \thisstar\ is less active than other Hyades stars in both activity indices, but especially so in Ca II.}}
   \label{activity}
\end{figure*}

One possibility for explaining the longer-period variability on \thisstar\ and others like EPIC 211974724 is that we view these stars nearly pole-on and the variability timescale is dominated by the spot evolution timescale/activity lifetime rather than the stellar rotation period. This interpretation is consistent with our upper limit on the projected rotational velocity of about 2 \kms. Interestingly, if true, this explanation would imply that the planet candidate, \thisplanet, has an orbit significantly misaligned from its host's spin axis. {\ron A pole-on viewing geometry could also potentially explain the lower spectroscopic activity indicators as well if fewer active regions are visible from our line of sight.}

Another more mundane possibility is that the long-period variability is instrumental in origin, and the true activity signal of \thisstar\ is undetectable in the presence of long-timescale instrumental systematics. We think this explanation is unlikely. While \Kepler\ and K2 data do exhibit long-term systematics due to differential velocity aberration, the morphology of the long-term signal in the \thisstar\ light curve does not match typical instrumental signals in K2 data. If the signal were instrumental, its amplitude would be unusually high for a star of this brightness. Additionally, the amplitude and morphology of the signal does not depend on the size or shape of the photometric aperture used to extract the light curve. The long-period signal is large enough that it should be detectable in ground-based observations which could clarify its origin.\footnote{The 35 day period detected on the Praesepe star EPIC 211974724 has already passed this test; the signal was detected both in K2 and ground-based observations separated by 5 years, effectively ruling out instrumental artifacts {\ron \citep{agueros, douglas17}}.}

\subsection{Recovering and Confirming the Planet Candidate}\label{massmeasurement}
Confirming \thisplanet\ and determining its orbital period with radial velocity follow-up will be quite challenging. We estimate a planet mass of about 6.5 $\pm$ 2 \mearth\ using the probabilistic mass-radius relationship from \citet{wolfgang}, which corresponds to an RV semiamplitude of about 1.0 $\pm$ 0.4 \ms. While some short-period\footnote{The long orbital period of \thisstar\ poses an additional challenge. Most advances in treating stellar activity signals have been for exoplanets with orbital periods shorter than the stellar rotation period \citep[e.g.][]{haywood2014}.} exoplanets with RV semiamplitudes this small have been detected, such small signals push against the limits of existing instrumentation and analysis techniques. Detecting such a small RV semiamplitude in the presence of the high-amplitude stellar activity signals expected for Hyades-age stars will be very difficult. Even in the optimistic case that \thisstar\ has an unusually slow rotation period of 37 days, given the amplitude of photometric variations observed during the K2 observations, we estimate the stellar activity would induce up to 6-8 \ms\ peak-to-peak RV variations. Detecting the smaller signal of \thisplanet\ in radial velocities may not be possible until instrumentation and analysis techniques have advanced. 

The most straightforward path to confirming the transit signal and precisely measuring the orbital period of \thisplanet\ is photometric monitoring to detect additional transits. The candidate's long orbital period and shallow depth make it infeasible to detect from the ground, so space-based monitoring is required. NASA's recently-launched Transiting Exoplanet Survey Satellite (TESS) mission \citep{ricker} will not observe \thisstar\ during its two-year prime mission because it lies too close to the ecliptic plane, but could observe \thisstar\ in an extended mission. In particular, some of the extended mission concepts proposed by \citet{bouma} observe the ecliptic plane for periods of time ranging from 14 days to up to 112 days. If one of these longer ecliptic pointings were to be adopted as a TESS extended mission, it could detect a transit of \thisplanet. The orbital period of the planet is probably just a bit longer than the 72 day minimum allowed orbital period, and TESS should be able to detect a transit of \thisplanet\ with a signal-to-noise ratio of about 11 \citep{ticgen, stassun}. The confirmation of a habitable-zone super-Earth in an open cluster would be a strong example of how K2-TESS synergy can strengthen the legacy of both missions. 

\acknowledgments
We thank Luke Bouma for helpful discussions about TESS extended mission strategies, and we thank the anonymous referee for a helpful and constructive review. This work was performed in part under contract with the California Institute of Technology/Jet Propulsion Laboratory funded by NASA through the Sagan Fellowship Program executed by the NASA Exoplanet Science Institute. AWM was supported through Hubble Fellowship grant 51364 awarded by the Space Telescope Science Institute, which is operated by the Association of Universities for Research in Astronomy, Inc., for NASA, under contract NAS 5-26555. ACR was supported (in part) by NASA K2 Guest Observer Cycle 4 grant NNX17AF71G. D.W.L. acknowledges partial support from the TESS mission through a sub-award from the Massachusetts Institute of Technology to the Smithsonian Astrophysical Observatory (SAO) and from the \Kepler\ mission under NASA Cooperative agreement NNX13AB58A with SAO.

This research has made use of NASA's Astrophysics Data System and the NASA Exoplanet Archive, which is operated by the California Institute of Technology, under contract with the National Aeronautics and Space Administration under the Exoplanet Exploration Program. The National Geographic Society--Palomar Observatory Sky Atlas (POSS-I) was made by the California Institute of Technology with grants from the National Geographic Society. The Oschin Schmidt Telescope is operated by the California Institute of Technology and Palomar Observatory.

This paper includes data collected by the \Kepler\ mission. Funding for the \Kepler\ mission is provided by the NASA Science Mission directorate. Some of the data presented in this paper were obtained from the Mikulski Archive for Space Telescopes (MAST). STScI is operated by the Association of Universities for Research in Astronomy, Inc., under NASA contract NAS5--26555. Support for MAST for non--HST data is provided by the NASA Office of Space Science via grant NNX13AC07G and by other grants and contracts. This work has made use of data from the European Space Agency (ESA) mission {\it Gaia} (\url{https://www.cosmos.esa.int/gaia}), processed by the {\it Gaia} Data Processing and Analysis Consortium (DPAC, \url{https://www.cosmos.esa.int/web/gaia/dpac/consortium}). Funding for the DPAC
has been provided by national institutions, in particular the institutions participating in the {\it Gaia} Multilateral Agreement.

Some observations in the paper made use of the NN-EXPLORE Exoplanet and Stellar Speckle Imager (NESSI). NESSI was funded by the NASA Exoplanet Exploration Program and the NASA Ames Research Center. NESSI was built at the Ames Research Center by Steve B. Howell, Nic Scott, Elliott P. Horch, and Emmett Quigley. The NESSI data were obtained at the WIYN Observatory from telescope time allocated to NN-EXPLORE through the scientific partnership of the National Aeronautics and Space Administration, the National Science Foundation, and the National Optical Astronomy Observatory.

We wish to recognize and acknowledge the very significant cultural role and reverence that the summit of Maunakea has always had within the indigenous Hawaiian community.  We are most fortunate to have the opportunity to conduct observations from this mountain. We are also honored to be permitted to conduct observations on Iolkam Du’ag (Kitt Peak), a mountain within the Tohono O'odham Nation with particular significance to the Tohono O'odham people.

Facilities: \facility{\Kepler/K2, FLWO:1.5m (TRES, CfA Digital Speedometers), WIYN (NESSI), Gemini:Gillett (\Alopeke), ORO:Wyeth (CfA Digital Speedometers), \Gaia, Exoplanet Archive, MAST, CDS, ADS}


\begin{thebibliography}{}
\expandafter\ifx\csname natexlab\endcsname\relax\def\natexlab#1{#1}\fi

\bibitem[{{Ag{\"u}eros} {et~al.}(2011){Ag{\"u}eros}, {Covey}, {Lemonias},
  {Law}, {Kraus}, {Batalha}, {Bloom}, {Cenko}, {Kasliwal}, {Kulkarni},
  {Nugent}, {Ofek}, {Poznanski}, \& {Quimby}}]{agueros}
{Ag{\"u}eros}, M.~A., {Covey}, K.~R., {Lemonias}, J.~J., {et~al.} 2011, \apj,
  740, 110

\bibitem[{{Allard} {et~al.}(2011){Allard}, {Homeier}, \&
  {Freytag}}]{Allard2011}
{Allard}, F., {Homeier}, D., \& {Freytag}, B. 2011, in Astronomical Society of
  the Pacific Conference Series, Vol. 448, 16th Cambridge Workshop on Cool
  Stars, Stellar Systems, and the Sun, ed. C.~{Johns-Krull}, M.~K. {Browning},
  \& A.~A. {West}, 91

\bibitem[{{Altmann} {et~al.}(2017){Altmann}, {Roeser}, {Demleitner}, {Bastian},
  \& {Schilbach}}]{hsoy}
{Altmann}, M., {Roeser}, S., {Demleitner}, M., {Bastian}, U., \& {Schilbach},
  E. 2017, \aap, 600, L4

\bibitem[{{Barnes}(2007)}]{barnes}
{Barnes}, S.~A. 2007, \apj, 669, 1167

\bibitem[{{Bessell} \& {Murphy}(2012)}]{2012PASP..124..140B}
{Bessell}, M., \& {Murphy}, S. 2012, \pasp, 124, 140

\bibitem[{{Boss}(1995)}]{boss}
{Boss}, A.~P. 1995, Science, 267, 360

\bibitem[{{Bouma} {et~al.}(2017){Bouma}, {Winn}, {Kosiarek}, \&
  {McCullough}}]{bouma}
{Bouma}, L.~G., {Winn}, J.~N., {Kosiarek}, J., \& {McCullough}, P.~R. 2017,
  ArXiv e-prints, arXiv:1705.08891

\bibitem[{{Brandt} \& {Huang}(2015)}]{Brandt2015}
{Brandt}, T.~D., \& {Huang}, C.~X. 2015, \apj, 807, 24

\bibitem[{{Buchhave} {et~al.}(2010){Buchhave}, {Bakos}, {Hartman}, {Torres},
  {Kov{\'a}cs}, {Latham}, {Noyes}, {Esquerdo}, {Everett}, {Howard}, {Marcy},
  {Fischer}, {Johnson}, {Andersen}, {F{\H u}r{\'e}sz}, {Perumpilly},
  {Sasselov}, {Stefanik}, {B{\'e}ky}, {L{\'a}z{\'a}r}, {Papp}, \&
  {S{\'a}ri}}]{buchhave10}
{Buchhave}, L.~A., {Bakos}, G.~{\'A}., {Hartman}, J.~D., {et~al.} 2010, \apj,
  720, 1118

\bibitem[{{Buchhave} {et~al.}(2012){Buchhave}, {Latham}, {Johansen},
  {Bizzarro}, {Torres}, {Rowe}, {Batalha}, {Borucki}, {Brugamyer}, {Caldwell},
  {Bryson}, {Ciardi}, {Cochran}, {Endl}, {Esquerdo}, {Ford}, {Geary},
  {Gilliland}, {Hansen}, {Isaacson}, {Laird}, {Lucas}, {Marcy}, {Morse},
  {Robertson}, {Shporer}, {Stefanik}, {Still}, \& {Quinn}}]{buchhave}
{Buchhave}, L.~A., {Latham}, D.~W., {Johansen}, A., {et~al.} 2012, \nat, 486,
  375

\bibitem[{{Buchhave} {et~al.}(2014){Buchhave}, {Bizzarro}, {Latham},
  {Sasselov}, {Cochran}, {Endl}, {Isaacson}, {Juncher}, \&
  {Marcy}}]{buchhave14}
{Buchhave}, L.~A., {Bizzarro}, M., {Latham}, D.~W., {et~al.} 2014, \nat, 509,
  593

\bibitem[{{Burke} {et~al.}(2006){Burke}, {Gaudi}, {DePoy}, \&
  {Pogge}}]{burkecluster}
{Burke}, C.~J., {Gaudi}, B.~S., {DePoy}, D.~L., \& {Pogge}, R.~W. 2006, \aj,
  132, 210

\bibitem[{{Butler} {et~al.}(1997){Butler}, {Marcy}, {Williams}, {Hauser}, \&
  {Shirts}}]{butler}
{Butler}, R.~P., {Marcy}, G.~W., {Williams}, E., {Hauser}, H., \& {Shirts}, P.
  1997, \apjl, 474, L115

\bibitem[{{Campbell} \& {Walker}(1979)}]{campbellwalker1}
{Campbell}, B., \& {Walker}, G.~A.~H. 1979, \pasp, 91, 540

\bibitem[{{Campbell} {et~al.}(1988){Campbell}, {Walker}, \&
  {Yang}}]{campbellwalker2}
{Campbell}, B., {Walker}, G.~A.~H., \& {Yang}, S. 1988, \apj, 331, 902

\bibitem[{{Charbonneau} {et~al.}(2009){Charbonneau}, {Berta}, {Irwin}, {Burke},
  {Nutzman}, {Buchhave}, {Lovis}, {Bonfils}, {Latham}, {Udry}, {Murray-Clay},
  {Holman}, {Falco}, {Winn}, {Queloz}, {Pepe}, {Mayor}, {Delfosse}, \&
  {Forveille}}]{charbonneau}
{Charbonneau}, D., {Berta}, Z.~K., {Irwin}, J., {et~al.} 2009, \nat, 462, 891

\bibitem[{{Choi} {et~al.}(2016){Choi}, {Dotter}, {Conroy}, {Cantiello},
  {Paxton}, \& {Johnson}}]{MIST1}
{Choi}, J., {Dotter}, A., {Conroy}, C., {et~al.} 2016, \apj, 823, 102

\bibitem[{{Ciardi} {et~al.}(2018){Ciardi}, {Crossfield}, {Feinstein},
  {Schlieder}, {Petigura}, {David}, {Bristow}, {Patel}, {Arnold}, {Benneke},
  {Christiansen}, {Dressing}, {Fulton}, {Howard}, {Isaacson}, {Sinukoff}, \&
  {Thackeray}}]{ciardi}
{Ciardi}, D.~R., {Crossfield}, I.~J.~M., {Feinstein}, A.~D., {et~al.} 2018,
  \aj, 155, 10

\bibitem[{{Claret} \& {Bloemen}(2011)}]{claretbloemen}
{Claret}, A., \& {Bloemen}, S. 2011, \aap, 529, A75

\bibitem[{{Cochran} {et~al.}(1997){Cochran}, {Hatzes}, {Butler}, \&
  {Marcy}}]{cochran97}
{Cochran}, W.~D., {Hatzes}, A.~P., {Butler}, R.~P., \& {Marcy}, G.~W. 1997,
  \apj, 483, 457

\bibitem[{{Cochran} {et~al.}(2002){Cochran}, {Hatzes}, \& {Paulson}}]{cochran}
{Cochran}, W.~D., {Hatzes}, A.~P., \& {Paulson}, D.~B. 2002, \aj, 124, 565

\bibitem[{{Cohen} {et~al.}(2003){Cohen}, {Wheaton}, \&
  {Megeath}}]{2003AJ....126.1090C}
{Cohen}, M., {Wheaton}, W.~A., \& {Megeath}, S.~T. 2003, \aj, 126, 1090

\bibitem[{{Curtis} {et~al.}(2018){Curtis}, {Vanderburg}, {Torres}, {Kraus},
  {Huber}, {Mann}, {Rizzuto}, {Isaacson}, {Howard}, {Henze}, {Fulton}, \&
  {Wright}}]{curtis}
{Curtis}, J.~L., {Vanderburg}, A., {Torres}, G., {et~al.} 2018, \aj, 155, 173

\bibitem[{{Cushing} {et~al.}(2005){Cushing}, {Rayner}, \&
  {Vacca}}]{Cushing:2005lr}
{Cushing}, M.~C., {Rayner}, J.~T., \& {Vacca}, W.~D. 2005, \apj, 623, 1115

\bibitem[{{David} {et~al.}(2016{\natexlab{a}}){David}, {Hillenbrand},
  {Petigura}, {Carpenter}, {Crossfield}, {Hinkley}, {Ciardi}, {Howard},
  {Isaacson}, {Cody}, {Schlieder}, {Beichman}, \& {Barenfeld}}]{david}
{David}, T.~J., {Hillenbrand}, L.~A., {Petigura}, E.~A., {et~al.}
  2016{\natexlab{a}}, \nat, 534, 658

\bibitem[{{David} {et~al.}(2016{\natexlab{b}}){David}, {Conroy}, {Hillenbrand},
  {Stassun}, {Stauffer}, {Rebull}, {Cody}, {Isaacson}, {Howard}, \&
  {Aigrain}}]{davidhyades}
{David}, T.~J., {Conroy}, K.~E., {Hillenbrand}, L.~A., {et~al.}
  2016{\natexlab{b}}, \aj, 151, 112

\bibitem[{{David} {et~al.}(2018){David}, {Mamajek}, {Vanderburg}, {Schlieder},
  {Bristow}, {Petigura}, {Ciardi}, {Crossfield}, {Isaacson}, {Cody},
  {Stauffer}, {Hillenbrand}, {Bieryla}, {Latham}, {Fulton}, {Rebull},
  {Beichman}, {Gonzales}, {Hirsch}, {Howard}, {Vasisht}, \& {Ygouf}}]{castau}
{David}, T.~J., {Mamajek}, E.~E., {Vanderburg}, A., {et~al.} 2018, ArXiv
  e-prints, arXiv:1801.07320

\bibitem[{{Dawson} {et~al.}(2015){Dawson}, {Murray-Clay}, \&
  {Johnson}}]{dawson}
{Dawson}, R.~I., {Murray-Clay}, R.~A., \& {Johnson}, J.~A. 2015, \apj, 798, 66

\bibitem[{{Dotter}(2016)}]{MIST0}
{Dotter}, A. 2016, \apjs, 222, 8

\bibitem[{{Dotter} {et~al.}(2008){Dotter}, {Chaboyer}, {Jevremovi{\'c}},
  {Kostov}, {Baron}, \& {Ferguson}}]{Dotter2008}
{Dotter}, A., {Chaboyer}, B., {Jevremovi{\'c}}, D., {et~al.} 2008, \apjs, 178,
  89

\bibitem[{{Douglas} {et~al.}(2016){Douglas}, {Ag{\"u}eros}, {Covey}, {Cargile},
  {Barclay}, {Cody}, {Howell}, \& {Kopytova}}]{douglas16}
{Douglas}, S.~T., {Ag{\"u}eros}, M.~A., {Covey}, K.~R., {et~al.} 2016, \apj,
  822, 47

\bibitem[{{Douglas} {et~al.}(2017){Douglas}, {Ag{\"u}eros}, {Covey}, \&
  {Kraus}}]{douglas17}
{Douglas}, S.~T., {Ag{\"u}eros}, M.~A., {Covey}, K.~R., \& {Kraus}, A. 2017,
  \apj, 842, 83

\bibitem[{{Douglas} {et~al.}(2014){Douglas}, {Ag{\"u}eros}, {Covey}, {Bowsher},
  {Bochanski}, {Cargile}, {Kraus}, {Law}, {Lemonias}, {Arce}, {Fierroz}, \&
  {Kundert}}]{douglas14}
{Douglas}, S.~T., {Ag{\"u}eros}, M.~A., {Covey}, K.~R., {et~al.} 2014, \apj,
  795, 161

\bibitem[{{Dutra-Ferreira} {et~al.}(2016){Dutra-Ferreira}, {Pasquini},
  {Smiljanic}, {Porto de Mello}, \& {Steffen}}]{dutraferreira}
{Dutra-Ferreira}, L., {Pasquini}, L., {Smiljanic}, R., {Porto de Mello}, G.~F.,
  \& {Steffen}, M. 2016, \aap, 585, A75

\bibitem[{{ESA}(1997)}]{esahipparcos}
{ESA}, ed. 1997, ESA Special Publication, Vol. 1200, {The HIPPARCOS and TYCHO
  catalogues. Astrometric and photometric star catalogues derived from the ESA
  HIPPARCOS Space Astrometry Mission}

\bibitem[{{Gaia Collaboration} {et~al.}(2018){Gaia Collaboration}, {Brown},
  {Vallenari}, {Prusti}, {de Bruijne}, {Babusiaux}, \&
  {Bailer-Jones}}]{gaiadr2}
{Gaia Collaboration}, {Brown}, A.~G.~A., {Vallenari}, A., {et~al.} 2018, ArXiv
  e-prints, arXiv:1804.09365

\bibitem[{{Gaia Collaboration} {et~al.}(2016{\natexlab{a}}){Gaia
  Collaboration}, {Brown}, {Vallenari}, {Prusti}, {de Bruijne}, {Mignard},
  {Drimmel}, {Babusiaux}, {Bailer-Jones}, {Bastian}, \& et~al.}]{gaiadr1}
---. 2016{\natexlab{a}}, \aap, 595, A2

\bibitem[{{Gaia Collaboration} {et~al.}(2016{\natexlab{b}}){Gaia
  Collaboration}, {Prusti}, {de Bruijne}, {Brown}, {Vallenari}, {Babusiaux},
  {Bailer-Jones}, {Bastian}, {Biermann}, {Evans}, \& et~al.}]{gaiamission}
{Gaia Collaboration}, {Prusti}, T., {de Bruijne}, J.~H.~J., {et~al.}
  2016{\natexlab{b}}, \aap, 595, A1

\bibitem[{{Gaidos} {et~al.}(2014){Gaidos}, {Mann}, {L{\'e}pine}, {Buccino},
  {James}, {Ansdell}, {Petrucci}, {Mauas}, \& {Hilton}}]{Gaidos2014}
{Gaidos}, E., {Mann}, A.~W., {L{\'e}pine}, S., {et~al.} 2014, \mnras, 443, 2561

\bibitem[{{Geller} \& {Mathieu}(2011)}]{geller}
{Geller}, A.~M., \& {Mathieu}, R.~D. 2011, \nat, 478, 356

\bibitem[{{Gilliland} {et~al.}(2000){Gilliland}, {Brown}, {Guhathakurta},
  {Sarajedini}, {Milone}, {Albrow}, {Baliber}, {Bruntt}, {Burrows},
  {Charbonneau}, {Choi}, {Cochran}, {Edmonds}, {Frandsen}, {Howell}, {Lin},
  {Marcy}, {Mayor}, {Naef}, {Sigurdsson}, {Stagg}, {Vandenberg}, {Vogt}, \&
  {Williams}}]{gilliland}
{Gilliland}, R.~L., {Brown}, T.~M., {Guhathakurta}, P., {et~al.} 2000, \apjl,
  545, L47

\bibitem[{Goodman \& Weare(2010)}]{goodman}
Goodman, J., \& Weare, J. 2010, Communications in Applied Mathematics and
  Computational Science, 5, 65

\bibitem[{{Griffin} {et~al.}(1988){Griffin}, {Griffin}, {Gunn}, \&
  {Zimmerman}}]{griffin}
{Griffin}, R.~F., {Griffin}, R.~E.~M., {Gunn}, J.~E., \& {Zimmerman}, B.~A.
  1988, \aj, 96, 172

\bibitem[{{Haywood} {et~al.}(2014){Haywood}, {Collier Cameron}, {Queloz},
  {Barros}, {Deleuil}, {Fares}, {Gillon}, {Lanza}, {Lovis}, {Moutou}, {Pepe},
  {Pollacco}, {Santerne}, {S{\'e}gransan}, \& {Unruh}}]{haywood2014}
{Haywood}, R.~D., {Collier Cameron}, A., {Queloz}, D., {et~al.} 2014, \mnras,
  443, 2517

\bibitem[{{Henden} {et~al.}(2012){Henden}, {Levine}, {Terrell}, {Smith}, \&
  {Welch}}]{Henden:2012fk}
{Henden}, A.~A., {Levine}, S.~E., {Terrell}, D., {Smith}, T.~C., \& {Welch}, D.
  2012, Journal of the American Association of Variable Star Observers
  (JAAVSO), 40, 430

\bibitem[{{Henry} \& {McCarthy}(1993)}]{Henry:1993fk}
{Henry}, T.~J., \& {McCarthy}, Jr., D.~W. 1993, \aj, 106, 773

\bibitem[{{H{\o}g} {et~al.}(2000){H{\o}g}, {Fabricius}, {Makarov}, {Urban},
  {Corbin}, {Wycoff}, {Bastian}, {Schwekendiek}, \& {Wicenec}}]{Hog2000}
{H{\o}g}, E., {Fabricius}, C., {Makarov}, V.~V., {et~al.} 2000, \aap, 355, L27

\bibitem[{{Howell} {et~al.}(2011){Howell}, {Everett}, {Sherry}, {Horch}, \&
  {Ciardi}}]{howell2011}
{Howell}, S.~B., {Everett}, M.~E., {Sherry}, W., {Horch}, E., \& {Ciardi},
  D.~R. 2011, \aj, 142, 19

\bibitem[{{Howell} {et~al.}(2014){Howell}, {Sobeck}, {Haas}, {Still},
  {Barclay}, {Mullally}, {Troeltzsch}, {Aigrain}, {Bryson}, {Caldwell},
  {Chaplin}, {Cochran}, {Huber}, {Marcy}, {Miglio}, {Najita}, {Smith},
  {Twicken}, \& {Fortney}}]{howell}
{Howell}, S.~B., {Sobeck}, C., {Haas}, M., {et~al.} 2014, \pasp, 126, 398

\bibitem[{{Huber} {et~al.}(2016){Huber}, {Bryson}, {Haas}, {Barclay},
  {Barentsen}, {Howell}, {Sharma}, {Stello}, \& {Thompson}}]{epic}
{Huber}, D., {Bryson}, S.~T., {Haas}, M.~R., {et~al.} 2016, \apjs, 224, 2

\bibitem[{{Jaffe} \& {Barclay}(2017)}]{ticgen}
{Jaffe}, T.~J., \& {Barclay}, T. 2017, ticgen: A tool for calculating a TESS
  magnitude, and an expected noise level for stars to be observed by TESS,
  doi:10.5281/zenodo.888217

\bibitem[{{Kipping}(2013)}]{kipping_prior1}
{Kipping}, D.~M. 2013, \mnras, 434, L51

\bibitem[{{Kipping}(2014)}]{kipping_prior2}
---. 2014, \mnras, 444, 2263

\bibitem[{{Kopparapu} {et~al.}(2013){Kopparapu}, {Ramirez}, {Kasting}, {Eymet},
  {Robinson}, {Mahadevan}, {Terrien}, {Domagal-Goldman}, {Meadows}, \&
  {Deshpande}}]{kopparapu}
{Kopparapu}, R.~K., {Ramirez}, R., {Kasting}, J.~F., {et~al.} 2013, \apj, 765,
  131

\bibitem[{{Kov{\'a}cs} {et~al.}(2002){Kov{\'a}cs}, {Zucker}, \&
  {Mazeh}}]{kovacs}
{Kov{\'a}cs}, G., {Zucker}, S., \& {Mazeh}, T. 2002, \aap, 391, 369

\bibitem[{{Lallement} {et~al.}(2003){Lallement}, {Welsh}, {Vergely}, {Crifo},
  \& {Sfeir}}]{Lallement_Bubble}
{Lallement}, R., {Welsh}, B.~Y., {Vergely}, J.~L., {Crifo}, F., \& {Sfeir}, D.
  2003, \aap, 411, 447

\bibitem[{{Latham} {et~al.}(1989){Latham}, {Stefanik}, {Mazeh}, {Mayor}, \&
  {Burki}}]{lathambd}
{Latham}, D.~W., {Stefanik}, R.~P., {Mazeh}, T., {Mayor}, M., \& {Burki}, G.
  1989, \nat, 339, 38

\bibitem[{{Leiner} {et~al.}(2016){Leiner}, {Mathieu}, {Stello}, {Vanderburg},
  \& {Sandquist}}]{leiner}
{Leiner}, E., {Mathieu}, R.~D., {Stello}, D., {Vanderburg}, A., \& {Sandquist},
  E. 2016, \apjl, 832, L13

\bibitem[{{Libralato} {et~al.}(2016){Libralato}, {Nardiello}, {Bedin},
  {Borsato}, {Granata}, {Malavolta}, {Piotto}, {Ochner}, {Cunial}, \&
  {Nascimbeni}}]{libralato}
{Libralato}, M., {Nardiello}, D., {Bedin}, L.~R., {et~al.} 2016, \mnras, 463,
  1780

\bibitem[{{Liu} {et~al.}(2016){Liu}, {Yong}, {Asplund}, {Ram{\'{\i}}rez}, \&
  {Mel{\'e}ndez}}]{Liu2016}
{Liu}, F., {Yong}, D., {Asplund}, M., {Ram{\'{\i}}rez}, I., \& {Mel{\'e}ndez},
  J. 2016, \mnras, 457, 3934

\bibitem[{{Livingston} {et~al.}(2018){Livingston}, {Dai}, {Hirano}, {Gandolfi},
  {Nowak}, {Endl}, {Velasco}, {Fukui}, {Narita}, {Prieto-Arranz}, {Barragan},
  {Cusano}, {Albrecht}, {Cabrera}, {Cochran}, {Csizmadia}, {Deeg},
  {Eigm{\"u}ller}, {Erikson}, {Fridlund}, {Grziwa}, {Guenther}, {Hatzes},
  {Kawauchi}, {Korth}, {Nespral}, {Palle}, {P{\"a}tzold}, {Persson}, {Rauer},
  {Smith}, {Tamura}, {Tanaka}, {Van Eylen}, {Watanabe}, \& {Winn}}]{livingston}
{Livingston}, J.~H., {Dai}, F., {Hirano}, T., {et~al.} 2018, \aj, 155, 115

\bibitem[{{Lopez} \& {Fortney}(2014)}]{lopezfortney14}
{Lopez}, E.~D., \& {Fortney}, J.~J. 2014, \apj, 792, 1

\bibitem[{{Lovis} \& {Mayor}(2007)}]{lovismayor}
{Lovis}, C., \& {Mayor}, M. 2007, \aap, 472, 657

\bibitem[{{Mandel} \& {Agol}(2002)}]{mandelagol}
{Mandel}, K., \& {Agol}, E. 2002, \apjl, 580, L171

\bibitem[{{Mann} {et~al.}(2015){Mann}, {Feiden}, {Gaidos}, {Boyajian}, \& {von
  Braun}}]{Mann2015b}
{Mann}, A.~W., {Feiden}, G.~A., {Gaidos}, E., {Boyajian}, T., \& {von Braun},
  K. 2015, \apj, 804, 64

\bibitem[{{Mann} \& {von Braun}(2015)}]{Mann2015a}
{Mann}, A.~W., \& {von Braun}, K. 2015, \pasp, 127, 102

\bibitem[{{Mann} {et~al.}(2016{\natexlab{a}}){Mann}, {Gaidos}, {Mace},
  {Johnson}, {Bowler}, {LaCourse}, {Jacobs}, {Vanderburg}, {Kraus}, {Kaplan},
  \& {Jaffe}}]{mann16}
{Mann}, A.~W., {Gaidos}, E., {Mace}, G.~N., {et~al.} 2016{\natexlab{a}}, \apj,
  818, 46

\bibitem[{{Mann} {et~al.}(2016{\natexlab{b}}){Mann}, {Newton}, {Rizzuto},
  {Irwin}, {Feiden}, {Gaidos}, {Mace}, {Kraus}, {James}, {Ansdell},
  {Charbonneau}, {Covey}, {Ireland}, {Jaffe}, {Johnson}, {Kidder}, \&
  {Vanderburg}}]{mann16b}
{Mann}, A.~W., {Newton}, E.~R., {Rizzuto}, A.~C., {et~al.} 2016{\natexlab{b}},
  \aj, 152, 61

\bibitem[{{Mann} {et~al.}(2017){Mann}, {Gaidos}, {Vanderburg}, {Rizzuto},
  {Ansdell}, {Medina}, {Mace}, {Kraus}, \& {Sokal}}]{Mann:2017aa}
{Mann}, A.~W., {Gaidos}, E., {Vanderburg}, A., {et~al.} 2017, \aj, 153, 64

\bibitem[{{Mann} {et~al.}(2018){Mann}, {Vanderburg}, {Rizzuto}, {Kraus},
  {Berlind}, {Bieryla}, {Calkins}, {Esquerdo}, {Latham}, {Mace}, {Morris},
  {Quinn}, {Sokal}, \& {Stefanik}}]{mann2018}
{Mann}, A.~W., {Vanderburg}, A., {Rizzuto}, A.~C., {et~al.} 2018, \aj, 155, 4

\bibitem[{{Mart{\'{\i}}n} {et~al.}(2018){Mart{\'{\i}}n}, {Lodieu}, {Pavlenko},
  \& {B{\'e}jar}}]{2018arXiv180207155M}
{Mart{\'{\i}}n}, E.~L., {Lodieu}, N., {Pavlenko}, Y., \& {B{\'e}jar}, V.~J.~S.
  2018, ArXiv e-prints

\bibitem[{{Masuda}(2014)}]{masuda51}
{Masuda}, K. 2014, \apj, 783, 53

\bibitem[{{Masuda} \& {Winn}(2017)}]{masudawinn}
{Masuda}, K., \& {Winn}, J.~N. 2017, \aj, 153, 187

\bibitem[{{Mayo} {et~al.}(2018){Mayo}, {Vanderburg}, {Latham}, {Bieryla},
  {Morton}, {Buchhave}, {Dressing}, {Beichman}, {Berlind}, {Calkins}, {Ciardi},
  {Crossfield}, {Esquerdo}, {Everett}, {Gonzales}, {Hirsch}, {Horch}, {Howard},
  {Howell}, {Livingston}, {Patel}, {Petigura}, {Schlieder}, {Scott}, {Schumer},
  {Sinukoff}, {Teske}, \& {Winters}}]{mayo}
{Mayo}, A.~W., {Vanderburg}, A., {Latham}, D.~W., {et~al.} 2018, ArXiv
  e-prints, arXiv:1802.05277

\bibitem[{{Mayor} \& {Queloz}(1995)}]{mayor}
{Mayor}, M., \& {Queloz}, D. 1995, \nat, 378, 355

\bibitem[{{Meibom} {et~al.}(2013){Meibom}, {Torres}, {Fressin}, {Latham},
  {Rowe}, {Ciardi}, {Bryson}, {Rogers}, {Henze}, {Janes}, {Barnes}, {Marcy},
  {Isaacson}, {Fischer}, {Howell}, {Horch}, {Jenkins}, {Schuler}, \&
  {Crepp}}]{meibom}
{Meibom}, S., {Torres}, G., {Fressin}, F., {et~al.} 2013, \nat, 499, 55

\bibitem[{{Mermilliod} {et~al.}(2009){Mermilliod}, {Mayor}, \&
  {Udry}}]{Mermilliod}
{Mermilliod}, J.-C., {Mayor}, M., \& {Udry}, S. 2009, \aap, 498, 949

\bibitem[{{Mermilliod} {et~al.}(1997){Mermilliod}, {Mermilliod}, \&
  {Hauck}}]{Mermilliod1997}
{Mermilliod}, J.-C., {Mermilliod}, M., \& {Hauck}, B. 1997, \aaps, 124, 349

\bibitem[{{Morton} {et~al.}(2016){Morton}, {Bryson}, {Coughlin}, {Rowe},
  {Ravichandran}, {Petigura}, {Haas}, \& {Batalha}}]{morton16}
{Morton}, T.~D., {Bryson}, S.~T., {Coughlin}, J.~L., {et~al.} 2016, \apj, 822,
  86

\bibitem[{{Mui{\~n}os} \& {Evans}(2014)}]{CMC15}
{Mui{\~n}os}, J.~L., \& {Evans}, D.~W. 2014, Astronomische Nachrichten, 335,
  367

\bibitem[{{M{\"u}ller} {et~al.}(2013){M{\"u}ller}, {Huber}, {Czesla}, {Wolter},
  \& {Schmitt}}]{muller}
{M{\"u}ller}, H.~M., {Huber}, K.~F., {Czesla}, S., {Wolter}, U., \& {Schmitt},
  J.~H.~M.~M. 2013, \aap, 560, A112

\bibitem[{{Naef} {et~al.}(2001){Naef}, {Latham}, {Mayor}, {Mazeh}, {Beuzit},
  {Drukier}, {Perrier-Bellet}, {Queloz}, {Sivan}, {Torres}, {Udry}, \&
  {Zucker}}]{hd80606}
{Naef}, D., {Latham}, D.~W., {Mayor}, M., {et~al.} 2001, \aap, 375, L27

\bibitem[{{Obermeier} {et~al.}(2016){Obermeier}, {Henning}, {Schlieder},
  {Crossfield}, {Petigura}, {Howard}, {Sinukoff}, {Isaacson}, {Ciardi},
  {David}, {Hillenbrand}, {Beichman}, {Howell}, {Horch}, {Everett}, {Hirsch},
  {Teske}, {Christiansen}, {L{\'e}pine}, {Aller}, {Liu}, {Saglia},
  {Livingston}, \& {Kluge}}]{obermeier}
{Obermeier}, C., {Henning}, T., {Schlieder}, J.~E., {et~al.} 2016, \aj, 152,
  223

\bibitem[{{Paulson} {et~al.}(2004){Paulson}, {Cochran}, \& {Hatzes}}]{paulson}
{Paulson}, D.~B., {Cochran}, W.~D., \& {Hatzes}, A.~P. 2004, \aj, 127, 3579

\bibitem[{{Paulson} {et~al.}(2003){Paulson}, {Sneden}, \&
  {Cochran}}]{paulson2003}
{Paulson}, D.~B., {Sneden}, C., \& {Cochran}, W.~D. 2003, \aj, 125, 3185

\bibitem[{{Pecaut} \& {Mamajek}(2013)}]{mamajek}
{Pecaut}, M.~J., \& {Mamajek}, E.~E. 2013, \apjs, 208, 9

\bibitem[{{Pepper} {et~al.}(2008){Pepper}, {Stanek}, {Pogge}, {Latham},
  {DePoy}, {Siverd}, {Poindexter}, \& {Sivakoff}}]{pepper}
{Pepper}, J., {Stanek}, K.~Z., {Pogge}, R.~W., {et~al.} 2008, \aj, 135, 907

\bibitem[{{Perryman} {et~al.}(1998){Perryman}, {Brown}, {Lebreton}, {Gomez},
  {Turon}, {Cayrel de Strobel}, {Mermilliod}, {Robichon}, {Kovalevsky}, \&
  {Crifo}}]{perryman}
{Perryman}, M.~A.~C., {Brown}, A.~G.~A., {Lebreton}, Y., {et~al.} 1998, \aap,
  331, 81

\bibitem[{{Quinn} {et~al.}(2012){Quinn}, {White}, {Latham}, {Buchhave},
  {Cantrell}, {Dahm}, {F{\H u}r{\'e}sz}, {Szentgyorgyi}, {Geary}, {Torres},
  {Bieryla}, {Berlind}, {Calkins}, {Esquerdo}, \& {Stefanik}}]{twobs}
{Quinn}, S.~N., {White}, R.~J., {Latham}, D.~W., {et~al.} 2012, \apjl, 756, L33

\bibitem[{{Rayner} {et~al.}(2009){Rayner}, {Cushing}, \& {Vacca}}]{Rayner2009}
{Rayner}, J.~T., {Cushing}, M.~C., \& {Vacca}, W.~D. 2009, \apjs, 185, 289

\bibitem[{{Rebull} {et~al.}(2017){Rebull}, {Stauffer}, {Hillenbrand}, {Cody},
  {Bouvier}, {Soderblom}, {Pinsonneault}, \& {Hebb}}]{rebull17}
{Rebull}, L.~M., {Stauffer}, J.~R., {Hillenbrand}, L.~A., {et~al.} 2017, \apj,
  839, 92

\bibitem[{{Ricker} {et~al.}(2015){Ricker}, {Winn}, {Vanderspek}, {Latham},
  {Bakos}, {Bean}, {Berta-Thompson}, {Brown}, {Buchhave}, {Butler}, {Butler},
  {Chaplin}, {Charbonneau}, {Christensen-Dalsgaard}, {Clampin}, {Deming},
  {Doty}, {De Lee}, {Dressing}, {Dunham}, {Endl}, {Fressin}, {Ge}, {Henning},
  {Holman}, {Howard}, {Ida}, {Jenkins}, {Jernigan}, {Johnson}, {Kaltenegger},
  {Kawai}, {Kjeldsen}, {Laughlin}, {Levine}, {Lin}, {Lissauer}, {MacQueen},
  {Marcy}, {McCullough}, {Morton}, {Narita}, {Paegert}, {Palle}, {Pepe},
  {Pepper}, {Quirrenbach}, {Rinehart}, {Sasselov}, {Sato}, {Seager},
  {Sozzetti}, {Stassun}, {Sullivan}, {Szentgyorgyi}, {Torres}, {Udry}, \&
  {Villasenor}}]{ricker}
{Ricker}, G.~R., {Winn}, J.~N., {Vanderspek}, R., {et~al.} 2015, Journal of
  Astronomical Telescopes, Instruments, and Systems, 1, 014003

\bibitem[{{Rizzuto} {et~al.}(2015){Rizzuto}, {Ireland}, \& {Kraus}}]{rizzuto15}
{Rizzuto}, A.~C., {Ireland}, M.~J., \& {Kraus}, A.~L. 2015, \mnras, 448, 2737

\bibitem[{{Rizzuto} {et~al.}(2011){Rizzuto}, {Ireland}, \&
  {Robertson}}]{rizzuto11}
{Rizzuto}, A.~C., {Ireland}, M.~J., \& {Robertson}, J.~G. 2011, \mnras, 416,
  3108

\bibitem[{{Rizzuto} {et~al.}(2017){Rizzuto}, {Mann}, {Vanderburg}, {Kraus}, \&
  {Covey}}]{rizzuto}
{Rizzuto}, A.~C., {Mann}, A.~W., {Vanderburg}, A., {Kraus}, A.~L., \& {Covey},
  K.~R. 2017, \aj, 154, 224

\bibitem[{{Rogers}(2015)}]{rogers}
{Rogers}, L.~A. 2015, \apj, 801, 41

\bibitem[{{R{\"o}ser} {et~al.}(2011){R{\"o}ser}, {Schilbach}, {Piskunov},
  {Kharchenko}, \& {Scholz}}]{roser}
{R{\"o}ser}, S., {Schilbach}, E., {Piskunov}, A.~E., {Kharchenko}, N.~V., \&
  {Scholz}, R.-D. 2011, \aap, 531, A92

\bibitem[{{Rowe} {et~al.}(2014){Rowe}, {Bryson}, {Marcy}, {Lissauer},
  {Jontof-Hutter}, {Mullally}, {Gilliland}, {Issacson}, {Ford}, {Howell},
  {Borucki}, {Haas}, {Huber}, {Steffen}, {Thompson}, {Quintana}, {Barclay},
  {Still}, {Fortney}, {Gautier}, {Hunter}, {Caldwell}, {Ciardi}, {Devore},
  {Cochran}, {Jenkins}, {Agol}, {Carter}, \& {Geary}}]{rowe}
{Rowe}, J.~F., {Bryson}, S.~T., {Marcy}, G.~W., {et~al.} 2014, \apj, 784, 45

\bibitem[{{Russell}(1914)}]{russell}
{Russell}, H.~N. 1914, Popular Astronomy, 22, 275

\bibitem[{{Sato} {et~al.}(2007){Sato}, {Izumiura}, {Toyota}, {Kambe}, {Takeda},
  {Masuda}, {Omiya}, {Murata}, {Itoh}, {Ando}, {Yoshida}, {Ikoma}, {Kokubo}, \&
  {Ida}}]{sato}
{Sato}, B., {Izumiura}, H., {Toyota}, E., {et~al.} 2007, \apj, 661, 527

\bibitem[{{Shapley}(1917)}]{shapley}
{Shapley}, H. 1917, \apj, 45, doi:10.1086/142314

\bibitem[{{Skrutskie} {et~al.}(2006){Skrutskie}, {Cutri}, {Stiening},
  {Weinberg}, {Schneider}, {Carpenter}, {Beichman}, {Capps}, {Chester},
  {Elias}, {Huchra}, {Liebert}, {Lonsdale}, {Monet}, {Price}, {Seitzer},
  {Jarrett}, {Kirkpatrick}, {Gizis}, {Howard}, {Evans}, {Fowler}, {Fullmer},
  {Hurt}, {Light}, {Kopan}, {Marsh}, {McCallon}, {Tam}, {Van Dyk}, \&
  {Wheelock}}]{Skrutskie2006}
{Skrutskie}, M.~F., {Cutri}, R.~M., {Stiening}, R., {et~al.} 2006, \aj, 131,
  1163

\bibitem[{{Stassun} {et~al.}(2017){Stassun}, {Oelkers}, {Pepper}, {Paegert},
  {De Lee}, {Torres}, {Latham}, {Muirhead}, {Dressing}, {Rojas-Ayala}, {Mann},
  {Fleming}, {Levine}, {Silvotti}, {Plavchan}, \& {the TESS Target Selection
  Working Group}}]{stassun}
{Stassun}, K.~G., {Oelkers}, R.~J., {Pepper}, J., {et~al.} 2017, ArXiv
  e-prints, arXiv:1706.00495

\bibitem[{{Stefanik} \& {Latham}(1985)}]{stefanik}
{Stefanik}, R.~P., \& {Latham}, D.~W. 1985, in Stellar Radial Velocities, ed.
  A.~G.~D. {Philip} \& D.~W. {Latham}, Vol.~88, 213--222

\bibitem[{{Stern} {et~al.}(1981){Stern}, {Zolcinski}, {Antiochos}, \&
  {Underwood}}]{stern}
{Stern}, R.~A., {Zolcinski}, M.~C., {Antiochos}, S.~K., \& {Underwood}, J.~H.
  1981, \apj, 249, 647

\bibitem[{{Struve}(1952)}]{struve}
{Struve}, O. 1952, The Observatory, 72, 199

\bibitem[{{Van Cleve} \& {Caldwell}(2016)}]{keplerinstrumenthandbook}
{Van Cleve}, J.~E., \& {Caldwell}, D.~A. 2016, {Kepler Instrument Handbook},
  Tech. rep.

\bibitem[{{van Leeuwen} {et~al.}(1997){van Leeuwen}, {Evans}, {Grenon},
  {Grossmann}, {Mignard}, \& {Perryman}}]{1997A&A...323L..61V}
{van Leeuwen}, F., {Evans}, D.~W., {Grenon}, M., {et~al.} 1997, \aap, 323, L61

\bibitem[{{van Saders} \& {Gaudi}(2011)}]{vansadersgaudi}
{van Saders}, J.~L., \& {Gaudi}, B.~S. 2011, \apj, 729, 63

\bibitem[{{Vanderburg}(2014)}]{v14}
{Vanderburg}, A. 2014, ArXiv e-prints, arXiv:1412.1827

\bibitem[{{Vanderburg} \& {Johnson}(2014)}]{vj14}
{Vanderburg}, A., \& {Johnson}, J.~A. 2014, \pasp, 126, 948

\bibitem[{{Vanderburg} {et~al.}(2015){Vanderburg}, {Montet}, {Johnson},
  {Buchhave}, {Zeng}, {Pepe}, {Collier Cameron}, {Latham}, {Molinari}, {Udry},
  {Lovis}, {Matthews}, {Cameron}, {Law}, {Bowler}, {Angus}, {Baranec},
  {Bieryla}, {Boschin}, {Charbonneau}, {Cosentino}, {Dumusque}, {Figueira},
  {Guenther}, {Harutyunyan}, {Hellier}, {Kuschnig}, {Lopez-Morales}, {Mayor},
  {Micela}, {Moffat}, {Pedani}, {Phillips}, {Piotto}, {Pollacco}, {Queloz},
  {Rice}, {Riddle}, {Rowe}, {Rucinski}, {Sasselov}, {S{\'e}gransan},
  {Sozzetti}, {Szentgyorgyi}, {Watson}, \& {Weiss}}]{hip116454}
{Vanderburg}, A., {Montet}, B.~T., {Johnson}, J.~A., {et~al.} 2015, \apj, 800,
  59

\bibitem[{{Vanderburg} {et~al.}(2016{\natexlab{a}}){Vanderburg}, {Becker},
  {Kristiansen}, {Bieryla}, {Duev}, {Jensen-Clem}, {Morton}, {Latham}, {Adams},
  {Baranec}, {Berlind}, {Calkins}, {Esquerdo}, {Kulkarni}, {Law}, {Riddle},
  {Salama}, \& {Schmitt}}]{hip41378}
{Vanderburg}, A., {Becker}, J.~C., {Kristiansen}, M.~H., {et~al.}
  2016{\natexlab{a}}, \apjl, 827, L10

\bibitem[{{Vanderburg} {et~al.}(2016{\natexlab{b}}){Vanderburg}, {Latham},
  {Buchhave}, {Bieryla}, {Berlind}, {Calkins}, {Esquerdo}, {Welsh}, \&
  {Johnson}}]{v15}
{Vanderburg}, A., {Latham}, D.~W., {Buchhave}, L.~A., {et~al.}
  2016{\natexlab{b}}, \apjs, 222, 14

\bibitem[{{Wolfgang} {et~al.}(2016){Wolfgang}, {Rogers}, \& {Ford}}]{wolfgang}
{Wolfgang}, A., {Rogers}, L.~A., \& {Ford}, E.~B. 2016, \apj, 825, 19

\bibitem[{{Wright} {et~al.}(2010){Wright}, {Eisenhardt}, {Mainzer}, {Ressler},
  {Cutri}, {Jarrett}, {Kirkpatrick}, {Padgett}, {McMillan}, {Skrutskie},
  {Stanford}, {Cohen}, {Walker}, {Mather}, {Leisawitz}, {Gautier}, {McLean},
  {Benford}, {Lonsdale}, {Blain}, {Mendez}, {Irace}, {Duval}, {Liu}, {Royer},
  {Heinrichsen}, {Howard}, {Shannon}, {Kendall}, {Walsh}, {Larsen}, {Cardon},
  {Schick}, {Schwalm}, {Abid}, {Fabinsky}, {Naes}, \& {Tsai}}]{Wright2010}
{Wright}, E.~L., {Eisenhardt}, P.~R.~M., {Mainzer}, A.~K., {et~al.} 2010, \aj,
  140, 1868

\end{thebibliography}

\begin{deluxetable*}{lcccc}
\tablecaption{System Parameters for \thisstar \label{bigtable}}
\tablewidth{0pt}
\tablehead{
  \colhead{Parameter} & 
  \colhead{Value}     &
  \colhead{} &
  \colhead{68.3\% Confidence}     &
  \colhead{Comment}   \\
  \colhead{} & 
  \colhead{}     &
  \colhead{} &
  \colhead{Interval Width}     &
  \colhead{}  
}
\startdata
\emph{Other Designations} & & & \\
EPIC 248045685  & & & \\
HIP 22271  & & & \\
BD+25 733 & & & \\
\\
\emph{Basic Information} & & & \\
Right Ascension & 04:47:41.80 & & & A \\
Declination & +26:09:00.8 & & & A \\
Proper Motion in RA [\ensuremath{\rm mas\,yr^{-1}}]& 113.42 & $\pm$ & 0.18&A  \\
Proper Motion in Dec [\ensuremath{\rm mas\,yr^{-1}}]& -83.83 & $\pm$ & 0.12&A  \\
Absolute Radial Velocity [\kms]& 39.84 & $\pm$ & 0.1& B \\
Distance to Star~[pc]& 47.51 &$\pm$&0.65& A\\
V-magnitude & 10.60 &$\pm$  & 0.012 & A\\ 
K-magnitude & 7.72 &$\pm$  & 0.03 & A\\ 
{\ron \Kepler-band Kp magnitude} & {\ron 10.15} & &  & {\ron A}\\
{\ron Mt. Wilson $R'_{\rm HK}$} & {\ron -4.77} & $\pm$ & 0.05 & {\ron B}\\
\\
\emph{Stellar Parameters} & & & \\
Mass $M_\star$~[$M_\odot$] & \mstar & $\pm$&$ \mstare$ & C \\
Radius $R_\star$~[$R_\odot$] & \rstar & $\pm$&$ \rstare$ & C \\
Luminosity $L_\star$~[$L_\odot$]& 0.182 & $\pm$& 0.006 & C \\
Limb darkening $u_1$~ & \ldone  & $\pm$&$ \uldone$ & D,E \\
Limb darkening $u_2$~ & \ldtwo  & $\pm$&$ \uldtwo$ & D,E \\
$\log g_{\rm SPC}$~[cgs] & \loggspc & $\pm$& \loggespc & B \\
Metallicity \metallicity & \mh & $\pm$&\mhe & F \\
$T_{\rm eff}$ [K] & 4655 & $\pm$& 55 & C\\
$v\sin{i}$ [\kms] &  $< 2$ & & & B\\

 & & \\
 
\emph{\thisplanet} & & & \\
Orbital Period, $P$~[days] & \perplb & &$ \uperplb $ & C,D \\
Radius Ratio, $R_P/R_\star$ & \rprstb & $\pm$ &$ \urprstb$ & C,D \\
Scaled semimajor axis, $a/R_\star$  & \arstb & &$ \uarstb$ & C,D \\
Orbital inclination, $i$~[deg] & \inclb & &$ \uinclb$ & C,D \\
Transit impact parameter, $b$ & \impb & &$ \uimpb$ & C,D \\
Transit Duration, $t_{14}$~[hours] & \tdurb &$\pm$ & \utdurb & D \\
Time of Transit $t_{t}$~[BJD] & \ttransitb & $\pm$& \uttransitb & D\\ 
Planet Radius $R_P$~[\rearth] & \rplb &   $\pm$& $\urplb$  & C,D \\
$T_{eq} = T_{\rm eff}(1 - \alpha)^{1/4}\sqrt{\frac{R_\star}{2a}}$~[K] & \teqb &   & $\uteqb$  & B,C,D,G \\
 & & \\

 & & \\

\enddata

\tablecomments{A: Parameters come from the EPIC catalog \citep{epic} and \Gaia\ Data Release 1 \citep{gaiadr1}. B: Parameters come from analysis of the three TRES spectra. C: Parameters come from measuring the bolometric flux and luminosity using archival photometry and the \Gaia\ parallax, and interpolating the measured stellar luminosity onto Hyades age isochrones as described (Section \ref{stellarparameters}). D: Parameters come from analysis of the K2 light curve (Section \ref{transitanalysis}) with priors on the orbital period imposed (Section \ref{orbperiod}). E: Gaussian priors of imposed on $u_1$ and $u_2$ centered on 0.644 and 0.096, respectively, with width 0.07.  F: The stellar metallicity is assumed to be the cluster metallicity. G: The equilibrium temperature $T_{eq}$ is calculated assuming albedo $\alpha$ uniformly distributed between 0 and 0.7 and perfect heat redistribution.  }

\end{deluxetable*}
\clearpage

\end{document}